\documentclass{JHEP3}

\usepackage{arydshln}

\usepackage{amsfonts,amssymb,amsmath,bm}
\usepackage{slashed}
\usepackage{comment}
\usepackage{graphicx}

\numberwithin{equation}{section}

    \newcommand{\AD}{AdS_5\times {\rm S}^5}
    
    \renewcommand{\[}{\left[}
    \renewcommand{\]}{\right]}
    \renewcommand{\(}{\left(}
    \renewcommand{\)}{\right)}

    \newcommand{\beq}{\begin{equation}}
    \newcommand{\eeq}{\end{equation}}
    \newcommand\beqa{\begin{eqnarray}}
    \newcommand\eeqa{\end{eqnarray}}

	\newcommand{\bea}{\begin{array}}
    \newcommand{\eea}{\end{array}}

    \newcommand{\nn}{\nonumber}

    \newcommand{\COMMENT}[1]{}
    
    \newcommand{\neqa}{\nonumber\end{eqnarray}}
    \newcommand{\la}[1]{\label{#1}}

    \newcommand{\eq}[1]{(\ref{#1})}

    \def\su2{{SU(2)}}

    \def\i2{\frac{i}{2}}

\def\s*{\ *_{\!\!\!\!\!\!\!\!\!\,_{\,_\text{\scriptsize{sym}}}}}
\def\hs*{\ \hat{*}_{\!\!\!\!\!\!\!\!\!\,_{\,_\text{\scriptsize{sym}}}}}

    \newcommand{\bg}{\begin{gather}}
    \newcommand{\eg}{\end{gather}}
    \newcommand{\bseq}{\begin{subequations}}
    \newcommand{\eseq}{\end{subequations}}

    \newcommand{\cO}{\mathcal{O}}

	\newcommand{\cY}{\mathcal{Y}}
	\newcommand{\cX}{\mathcal{X}}

	\newcommand{\cT}{\mathcal{T}}
	
	\newcommand{\cN}{\mathcal{N}}
	\newcommand{\cG}{\mathcal{G}}

	\newcommand{\bbC}{\mathbb{C}}
	
	\newcommand{\bQ}{{\bf Q}}
	\newcommand{\bT}{{\bf T}}

	\renewcommand{\Im}{{\rm Im}}
	
	\newcommand{\ofrac}[1]{\frac{1}{#1}}

\title{
Analytic Solution of Bremsstrahlung TBA II:
Turning on the Sphere Angle
}

\author{Nikolay Gromov$^{1,2}$,
  Fedor Levkovich-Maslyuk$^{1}$ and
  Grigory Sizov$^{1}$ \\
  $^1$King's College London, Department of Mathematics, \\ The Strand, London WC2R 2LS,
  United Kingdom\\\\
  $^2$ St.Petersburg INP, Gatchina, 188 300, St.Petersburg, Russia
  \qquad\\\\
  \textit{E-mail:}
  \email{nikgromov@gmail.com}, \email{fedor.levkovich@gmail.com}, \email{grigory.sizov@kcl.ac.uk }\\}

\abstract{
We find an exact analytical solution of
  the Y-system describing a cusped Wilson line
 in the planar limit of N=4 SYM.
 Our explicit solution describes anomalous dimensions of this family of observables
 for any value of the `t Hooft coupling
 and arbitrary R-charge $L$ of the local operator inserted on the cusp in a near-BPS limit.

 \hspace{5mm}
 Our finding generalizes the previous results of one of the authors \& Sever and
 passes several nontrivial tests.
First, for a particular case $L=0$ we reproduce the predictions of localization techniques.
Second, we show that in the classical limit our result perfectly reproduces the existing prediction from classical string theory.
In addition, we made a comparison with all existing weak coupling results and we found
that our result interpolates smoothly between these two very different regimes of AdS/CFT.
As a byproduct we found a generalization of the essential parts of the FiNLIE construction for the $\gamma$-deformed case
and discuss our results in the framework of the novel ${\bf P}\mu$-formulation of the spectral problem.
}

\preprint{}

\begin{document}


\section{Introduction}

The duality between planar $\cN=4$ supersymmetric Yang-Mills theory (SYM) in four dimensions and superstring theory in $\AD$ is one of the best-understood cases of AdS/CFT correspondence \cite{AdSCFT}. Recent intensive studies of this example have led to the discovery of integrable structures which
give a hope that the
 exact solution of both theories is within reach (for a review see e.g. \cite{Beisert:2010jr}).
Integrability-based methods have been used for observables such as correlation functions, Wilson loops and scattering amplitudes, and were especially successful in application to the spectral problem, providing a complete solution for the spectrum of anomalous dimensions of local single-trace operators in $\cN=4$ SYM.
The key ingredient of the integrability
structure is given by a nice universal set of functional relations (known as Y-system) \cite{Gromov:2009tv}
which together with the symmetry and analyticity constraints \cite{Cavaglia:2010nm,Gromov:2011cx}
can be related with an infinite set
of the Thermodynamic Bethe ansatz
integral equations \cite{Bombardelli:2009ns,Gromov:2009bc,Arutyunov:2009ur}.
The simple structure of Y-system and the underlying integrable Hirota equations
allows one \cite{Gromov:2011cx} to simplify considerably this infinite set of equations making possible
efficient perturbative expansion \cite{Leurent:2013mr} and high precision numerics
as well as the exact analytical derivations presented in this paper.

The Y-system approach was recently shown to be essential in understanding another kind of observable -- the quark-antiquark potential on the three-sphere, or equivalently the generalized cusp anomalous dimension $\Gamma_\text{cusp}$. This quantity describes the divergence in the expectation value of a Wilson loop made of two lines forming a cusp,
\beq
	\left\langle W\right\rangle\sim\left(\frac{\Lambda_{IR}}{\Lambda_{UV}}\right)^{\Gamma_\text{cusp}},
\eeq
with $\Lambda_{UV}$ and $\Lambda_{IR}$ being the UV and IR cutoffs \cite{PolyakovCusp}. The quantity $\Gamma_\text{cusp}$ has been studied at weak and strong coupling (for some recent results see \cite{Drukker:2011za, Correa:2012nkNew, BykZar, Henn:2013wfa}), and is also related to a number of other observables, such as IR divergence in amplitudes and radiation power from a moving quark, see e.g. \cite{Correa:2012at,Fiol:2012sg,Fiol:2013iaa,Kruczenski:2012aw}.
The cusp anomalous dimension is a function of two angles, $\phi$ and $\theta$, which describe the geometry of the Wilson line setup shown in Fig. \ref{pic:WLcusp} \cite{Drukker:1999zq}. The first angle, $\phi$, is the angle between the quark and antiquark lines at the cusp. The second angle, $\theta$, arises because the locally supersymmetric Wilson lines considered here include a coupling to the scalar fields.
 As there are six real scalars  in $\cN=4$ SYM the coupling can be defined by a unit vector $\vec n$ which gives a point on $S^5$. For the two lines we have two different vectors, $\vec n$ and $\vec n_\theta$, with $\theta$ being the angle between them. Explicitly, we can write the cusped Wilson loop as
\beq
W_0={\rm P}\exp\!\int\limits_{-\infty}^0\! dt\[i  A\cdot\dot{x}_q+\vec\Phi\cdot\vec n\,|\dot x_q|\]\times {\rm P}\exp\!\int\limits_0^\infty\!dt\[i A\cdot\dot x_{\bar q}+\vec\Phi\cdot\vec n_{\theta}\,|\dot x_{\bar q}|\],
\eeq
where $\vec \Phi$ denotes a vector consisting of the six scalars of $\cN=4$ SYM, while $x_q(t)$ and $x_{\bar q}(t)$ are the quark and antiquark trajectories (straight lines through the origin) which make up an angle $\phi$ at the cusp
(see Fig.\ref{pic:WLcusp}). 

\FIGURE[ht]
{\label{pic:WLcusp}

    \begin{tabular}{cc}
    \includegraphics[scale=0.3]{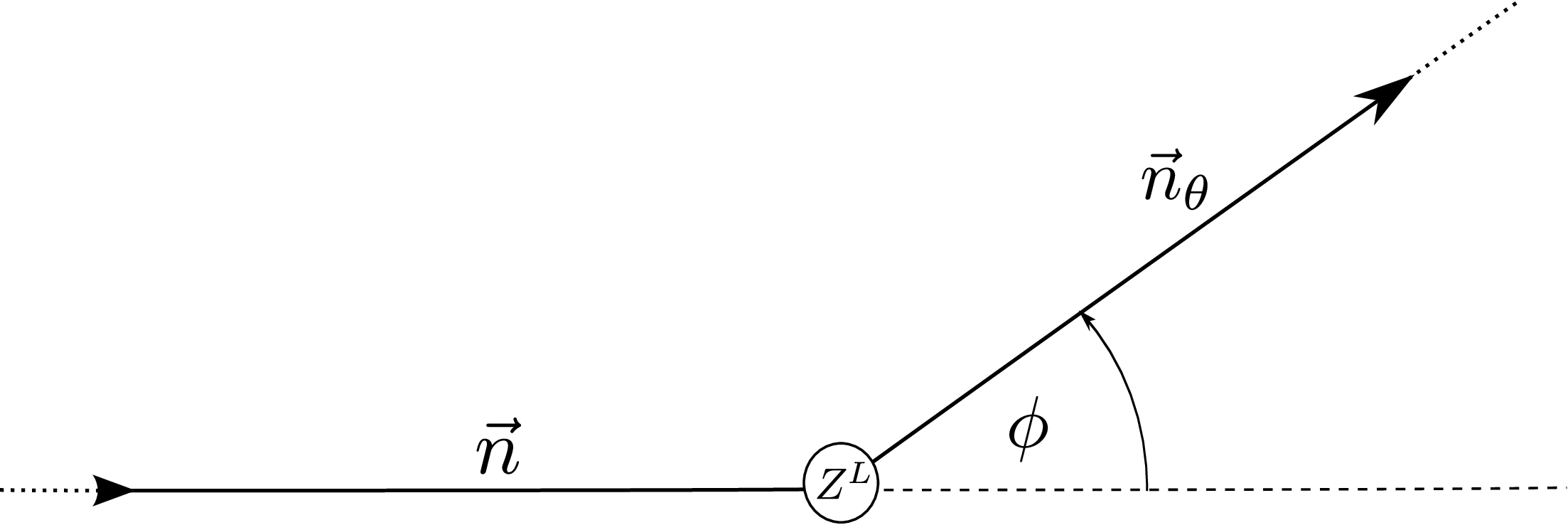}\\
    \end{tabular}
    \caption{\textbf{The setup.} A Wilson line with a cusp angle $\phi$ and $L$ scalar fields $Z=\Phi_1+i \Phi_2$ inserted at the cusp. Coupling of the scalar fields to the two half lines is defined by directions $\vec n$ and $\vec n_\theta$ in the internal space, with the angle $\theta$ between them.
    In this paper we consider the near-BPS limit corresponding to $\phi\approx\theta$. }
    }

A fully nonperturbative description for the value of $\Gamma_\text{cusp}$ was obtained in a remarkable development
by Drukker \cite{Drukker:2012de}
and
by Correa, Maldacena \& Sever \cite{Correa:2012hh}.
They proposed an infinite system of TBA integral equations which compute this quantity at arbitrary 't Hooft coupling $\lambda$ and for arbitrary angles. In order to implement the TBA approach, the cusp anomalous dimension was generalized for the case when a local operator with R-charge $L$ is inserted at the cusp (cf. Fig. \ref{pic:WLcusp}):
\beq
\label{WilsL}
	W_L={\rm P}\exp\!\int\limits_{-\infty}^0\! dt\(i  A\cdot\dot{x}_q+\vec\Phi\cdot\vec n\,|\dot x_q|\)\times Z^L\times {\rm P}\exp\!\int\limits_0^\infty\!dt\(i A\cdot\dot x_{\bar q}+\vec\Phi\cdot\vec n_\theta\,|\dot x_{\bar q}|\).
\eeq
Here $Z=\Phi_1+i\Phi_2$, with $\Phi_1$ and $\Phi_2$ being two scalars independent from $(\vec\Phi\cdot\vec n)$ and $(\vec\Phi\cdot\vec n_\theta)$. The anomalous dimension $\Gamma_L(\phi,\theta,\lambda)$ corresponding to such Wilson loop is captured by the TBA equations exactly at any value of $L$. For $L=0$ the usual quark-antiquark potential is recovered. The number of field insertions plays the role of the system's
volume in the TBA description, and $\Gamma_L(\phi,\theta,\lambda)$ is obtained as the vacuum state energy.

While the infinite system of these TBA equations is rather complicated, having the two angles as continuous parameters opens the possibility to look for simplifications in some limits
where an exact analytical solution may be expected\footnote{On the other hand, non-perturbative predictions from the spectral TBA have been mostly restricted to numerics \cite{Gromov:2009zb,Frolov:2010wt,Gromov:2011de,Frolov:2012zv}; see also \cite{Arutyunov:2012tx}.}.
In this paper we will focus on the near-BPS limit when $\phi\approx\theta$. For $\phi=\theta$ the configuration is BPS and the anomalous dimension vanishes \cite{Zarembo:2002an,DrukkerKawamoto}\footnote{Strictly speaking the BPS condition allows $\phi=-\theta$ in addition to $\phi=\theta$ but these two cases are trivially related.}. The small deviations from this supersymmetric case are known to be partially under control: the cusp dimension at $L=0$ was computed for $\phi\approx\theta$ analytically at any coupling in \cite{Correa:2012at,Fiol:2012sg} using results from localization methods \cite{localizations,Pestun}. The answer in the planar limit reads
\beq
\label{gammabps}
	\Gamma_\text{cusp}(\phi,\theta,\lambda)=-\ofrac{4\pi^2}(\phi^2-\theta^2)\ofrac{1-\frac{\theta^2}{\pi^2}}
	\frac{\sqrt{\tilde\lambda}\;I_2\(\sqrt{\tilde\lambda}\)}{I_1\(\sqrt{\tilde\lambda}\)}+\cO\((\phi^2-\theta^2)^2\),
	\ \  \tilde\lambda=\lambda\(1-\frac{\theta^2}{\pi^2}\)
\eeq
where $I_n$ are the modified Bessel functions of the first kind. The existence of such explicit result suggests that the cusp TBA
system should simplify dramatically when $\phi\approx\theta$.
Even though the full set of TBA equations was simplified a bit in this limit
as described in \cite{Correa:2012hh},
the result is still an enormously complicated infinite set of integral equations. Remarkably, it turned out that these equations admit an exact {\it analytical} solution. It was obtained in \cite{Gromov:2012eu} for the particular near-BPS configuration where $\theta=0$ and $\phi$ is small. The result of \cite{Gromov:2012eu} covers all values of $L$ and $\lambda$ and for $L=0$ reproduces the localization result \eq{gammabps} in which $\theta$ should be set to zero.

In the present paper we extend the results of \cite{Gromov:2012eu} to the generic near-BPS limit. Thus, we consider the case when $\phi\approx\theta$, but $\theta$ is arbitrary and is an extra parameter in the result.
We also filled some gaps in the previous derivation using the novel ${\bf P}\mu$-formulation \cite{PmuPRL}.
We obtain an explicit expression valid for all values of $\theta, \ L$ and $\lambda$. For this we solve the Bremsstrahlung TBA analytically, following the strategy developed in \cite{Gromov:2012eu}.
Quite surprisingly the result for arbitrary $\theta$ is considerably simpler and takes the form
\beq
\Gamma_L(g)=\frac{\phi-\theta}{4}\partial_\theta\log\frac{\det{\cal M}_{2L+1}}{\det{\cal M}_{2L-1}},\label{eq:mainresultIntro}
\eeq
where we define an $N+1\times N+1$ matrix

\beq
{\cal M}_{N}=\begin{pmatrix}
I_1^{\theta}& I_0^{\theta}& \cdots & I_{2-N}^{\theta}  &I_{1-N}^{\theta}\\
I_2^{\theta}& I_1^{\theta}& \cdots & I_{3-N}^{\theta} &I_{2-N}^{\theta}\\
\vdots      &  \vdots     &\ddots & \vdots            &\vdots           \\
I_{N}^{\theta}& I_{N-1}^{\theta}& \cdots & I_{1}^{\theta}  &I_{0}^{\theta}\\
I_{N+1}^{\theta}& I_{N}^{\theta}& \cdots & I_{2}^{\theta} &I_{1}^{\theta}\\
\end{pmatrix}
\eeq
and $I_n^\theta$ are
\beq
	I_n^\theta=\frac{1}{2}I_{n}\(\sqrt{\tilde\lambda}\)\[
	\(\sqrt{\frac{\pi+\theta}{\pi-\theta}}\)^{n}-
	(-1)^n\(\sqrt{\frac{\pi-\theta}{\pi+\theta}}\)^{n}
	\]. 
\eeq

At $L=0$ we have reproduced in full the localization result \eq{gammabps}. For $L>0$ our result complements and generalizes the calculation of \cite{Gromov:2012eu} as another integrability-based prediction for localization techniques. As in \cite{Gromov:2012eu}, the determinant expressions we got suggest a possible link to matrix models, which would be interesting to explore further.

The rest of the paper is organized as follows. In section \ref{sec:TBA} we describe the initial simplification of the TBA system in the near-BPS limit, resulting in an infinite set of the Bremsstrahlung TBA equations. Then in section \ref{sec:FiNLIE} we apply the powerful methods developed for the spectral problem to reduce this system to a finite set of equations, known as FiNLIE \cite{Gromov:2011cx,talk}\footnote{See
also \cite{seealso,seealso2} for an alternative approach.}. In section \ref{sec:Solving} we make an analytic ansatz for the unknowns in the FiNLIE and construct its explicit solution, obtaining our result for the energy. As in \cite{Gromov:2012eu} a key structure we encounter in the process is a Baxter equation for a set of auxiliary Bethe roots. We also describe checks of our result at both strong and weak coupling. In section \ref{sec:Pmu} we use our analytic solution of FiNLIE to illustrate a very recent reduction of the TBA equations to the so-called
${\bf P}\mu$-system \cite{PmuPRL} which involves only a few unknowns with simple analytical properties. In section \ref{sec:Conclusions} we present our conclusions. The several appendices contain various technical details.

\section{TBA equations in the near-BPS limit}
\label{sec:TBA}
In this section we discuss the first simplification of the cusp TBA system in the near-BPS regime, when
the two angles $\phi$ and $\theta$ are close to each other. Following \cite{Correa:2012hh} we will thus obtain a somewhat simpler, but still infinite, set of integral equations -- the Bremsstrahlung TBA.\footnote{The authors of \cite{Correa:2012hh} obtained the Bremsstrahlung TBA equations for the generic case $\phi\approx\theta$, but the equations were given explicitly in \cite{Correa:2012hh} only for the small angles case so we will repeat the derivation here.}

Let us remind that the cusp TBA equations are very similar to those describing the spectrum of single trace operator anomalous dimensions. After subtracting the asymptotic large $L$ solution, these two infinite sets of equations for the Y-functions $Y_{a,s}(u)$ become exactly the same. The integer indices $(a,s)$ of the Y-functions take values in the infinite T-shaped domain familiar from the spectral TBA (see Fig. \ref{pic:Yhook}). The only difference is in an extra symmetry requirement for the Y-functions, and in the large $L$ asymptotic solution\footnote{The extra symmetry requirement in the cusp TBA reads $Y_{a,s}(u)=Y_{a,-s}(-u)$ but is irrelevant in our discussion as for our state all Y-functions are even.}.

\FIGURE[ht]
{
\label{pic:Yhook}

    \begin{tabular}{cc}
    \includegraphics[scale=0.2]{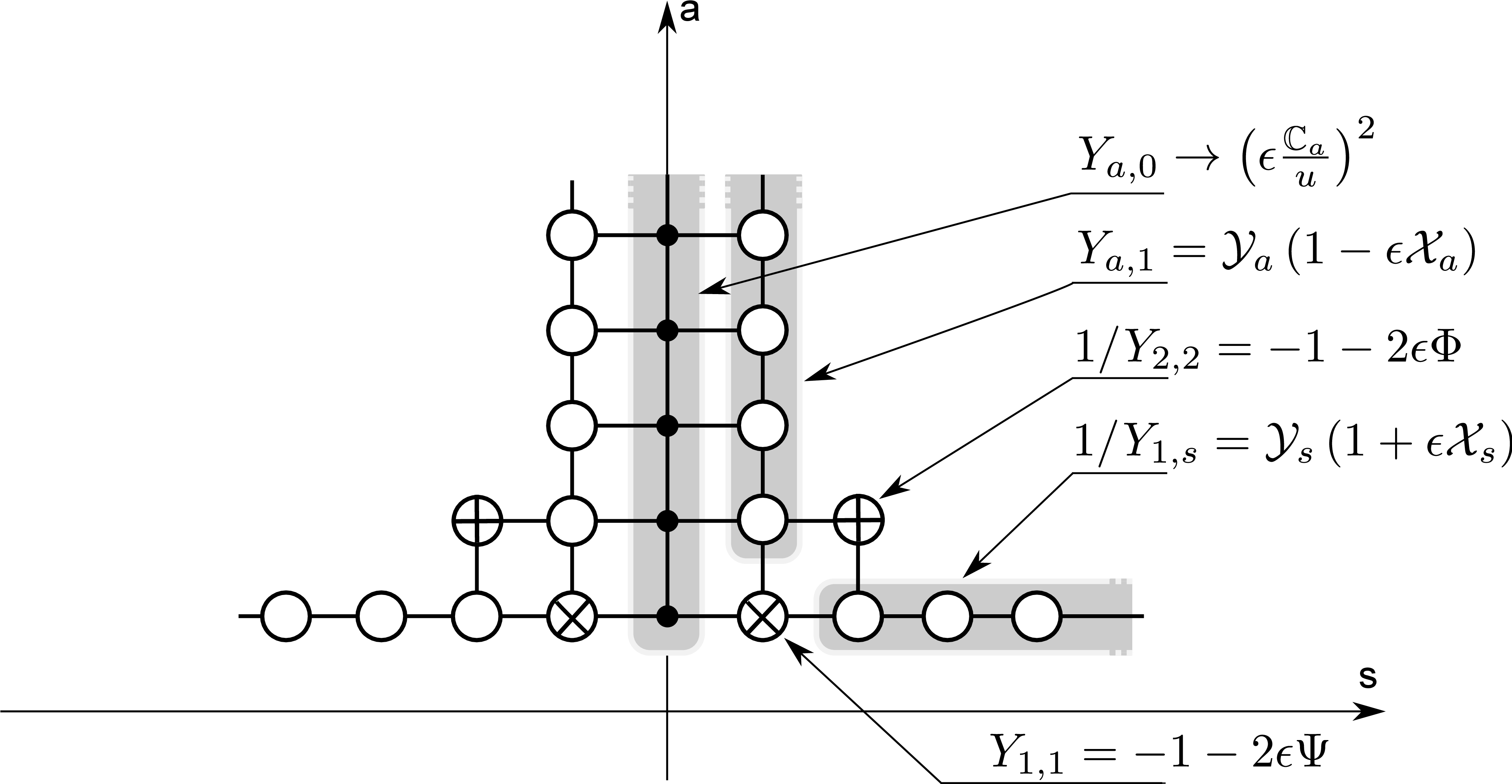}\\
    \end{tabular}
\caption{\textbf{The T-hook.} The indices $(a,s)$ of $Y$-functions take values on the infinite T-shaped lattice in the figure. We also show the form of expansion in small $\epsilon$ for different groups of $Y$-functions. Notice that the momentum carrying $Y$-functions $Y_{a,0}$ are small in $\epsilon$ and enter the system only through the singularity at $u=0$.}
}

The asymptotic solution encodes, in particular, the boundary scattering phase which has a double pole at zero mirror momentum. Due to this, the momentum-carrying functions $Y_{a,0}(u)$ have a double pole for $u=0$. This greatly simplifies their dynamics in the near-BPS regime -- only the residue at this pole is important and gives a non-vanishing contribution. This residue is small for $\phi\approx\theta$, and thus the structure of the expansion of the cusp TBA system in our case is very similar to what happens in the small angles regime discussed in detail in \cite{Correa:2012hh,Gromov:2012eu}.

We found it convenient to use a small expansion parameter
\beq
	\epsilon\equiv(\phi-\theta)\tan\phi_0,
\eeq
where\footnote{To shorten notation we will sometimes use $\theta$
instead of $\phi_0$ in the text, on the understanding that equations containing $\theta$ are assumed to hold to the leading order in $\epsilon$.} we denote $\phi_0=(\phi+\theta)/2$. As in the small angles case, it is sufficient to keep only the leading orders in the expansion of the Y-functions, which are
\beqa
\label{yexp1}
	&&Y_{a,1}=\cY_a\[1+\epsilon(\Omega_a-\cX_a)\],\ 1/Y_{1,s}=\cY_s\[1+\epsilon(\Omega_s+\cX_s)\],\
	\\ \nn
	&&Y_{1,1}=-1-2\epsilon\Psi,\ 1/Y_{2,2}=-1-2\epsilon\Phi,
\eeqa
while the residue of $Y_{a,0}$ reads
\beq
	\lim_{u\to0}\(u^2Y_{a,0}\)=\(\epsilon\,{\mathbb C}_a\)^2.
\eeq
This expansion (except for the $\Omega_a$ functions which will not enter our equations) is also shown in Fig. \ref{pic:Yhook}.

It is straightforward to plug these expansions into the cusp TBA system, and then simplify the equations a bit further using the same techniques as in the small angles case. We give more technical details in Appendix \ref{sec:BremsstrahlungApp}. The resulting set of Bremsstrahlung TBA equations reads:
\begin{align}
\label{eq:TBAfinaleq1}
	&\Phi-\Psi=\pi \mathbb{C}_a \hat K_{a}(u), \\
\label{eq:TBAfinaleq2}
	&\Phi+\Psi=\mathbf{s}*\left[-2\frac{{\cal X}_2}{1+{\cal Y}_2}+\pi(\hat K_a^+-\hat K_a^-)\mathbb{C}_a-\pi \delta(u) \mathbb{C}_1\right],
\\
\label{eq:TBAfinaleq3}
&\log Y_{1,m}=\mathbf{s}*I_{m,n}\log\left(1+Y_{1,n}\right)-\delta_{m,2}\textbf{s}\hat{*}\left(\log\frac{\Phi}{\Psi}+\epsilon\left(\Phi-\Psi\right)\right)-\epsilon \pi \mathbf{s} \mathbb{C}_m,
\\
\label{eq:TBAfinaleq5}
	&\Delta_a=[{\cal R}_{ab}^{(10)}+{\cal B}_{a,b-2}^{(10)}]\hat *\log\frac{1+{\cal Y}_b}{1+A_b}+{\cal R}_{a1}^{(10)}\hat *\log\left(\frac{\Psi}{1/2}\right)-
	{\cal B}_{a1}^{(10)}\hat *\log\left(\frac{\Phi}{1/2}\right),
\\
\label{eq:TBAfinaleq6}
	&\mathbb{C}_a=(-1)^{a+1} a \frac{\sin a\theta}{\tan\theta}\left(\sqrt{1+\frac{a^2}{16g^2}}-\frac{a}{4g}\right)^{2+2L}  F(a,g)e^{\Delta_a},
\end{align}
where the kernels and conventions are the same as in \cite{Gromov:2012eu} and are defined in Appendix \ref{sec:notationandconventions} -- in particular we use the standard notation
\beq
 f^{[\pm a]}\equiv f(u\pm ia),\ \ f^{\pm}\equiv f(u\pm i/2).
\eeq
The equation \eq{eq:TBAfinaleq3} for $Y_{1,m}$ should be understood to hold at orders $\cO(\epsilon^0)$ and $\cO(\epsilon^1)$ only. Notice that as in the small angles case
the functions $\Omega_a$ from \eq{yexp1} have dropped out of the equations.

We see that our Bremsstrahlung TBA equations are almost the same as in \cite{Gromov:2012eu}. However, importantly, the asymptotic condition at large real $u$ is different:
\beq
\label{bbYas}
	1/Y_{1,m}\to
\frac{\sin^2\theta}{\sin(m+1)\theta\sin(m-1)\theta},
\eeq
which should hold up to terms of order $\cO(\epsilon)$ inclusive.
Finally, the cusp anomalous dimension is determined by the double pole of momentum-carrying $Y$-functions:

\beq
\Gamma_L(g)=\epsilon\sum_{a=1}^\infty\frac{{\mathbb C}_a}{\sqrt{1+16g^2/a^2}}.
\label{eq:Gamma}
\eeq

In the next section we will reduce this TBA system to a finite set of nonlinear equations.

\section{FiNLIE}
\label{sec:FiNLIE}
\subsection{Twisted ansatz for T-functions}
In this section we apply the same methods as in \cite{Gromov:2012eu} to reduce the Bremsstrahlung TBA given above to a finite set of nonlinear integral equations
(FiNLIE). The FiNLIE approach of \cite{Gromov:2012eu} is very helpful to truly reveal the power of the spectral TBA
\cite{Leurent:2012ab, Leurent:2013mr}
\footnote{One can also use the Luscher approximation to extract the first several orders like in \cite{Bajnok:2008qj,Bajnok:2009vm,Bajnok:2012bz}.}.
For us it
allows to reduce drastically the number of unknown functions, opening the way to the analytic solution of the problem in section \ref{sec:Solving}.

Our main task is to reduce the infinite set of equations \eq{eq:TBAfinaleq3} for the functions $Y_{1,m}$. In order to do this we use its relation to the Y-system and Hirota equations in the horizontal right wing of the T-hook. Indeed,
from the integral form of \eq{eq:TBAfinaleq3} and the analyticity of the kernels it is clear that $Y_{1,m}(u)$ are analytic and regular in the strip $|\Im \, u|<\frac{m-1}{2}$. Then for $m>2$ the equation \eq{eq:TBAfinaleq3} can be rewritten as the Y-system functional equation using the property \eqref{eq:invtranslationdeformed}:
\beq\la{Y1m}
	\log\(Y_{1,m}^+Y_{1,m}^-\)=\log\(1+Y_{1,m-1}\)\(1+Y_{1,m+1}\).
\eeq

This set of functional equations can be solved by switching to the so-called T-functions according to
\beq
\label{YthroughTfull}
1/{Y}_{1,m}=\frac{{T}_{1,m}^+ {T}_{1,m}^-}{{T}_{1,m+1} {T}_{1,m-1}}-1.
\eeq

In terms of T-functions the Y-system equation becomes the Hirota equation in the horizontal strip, for which the general solution is known
\cite{Gromov:2011cx,Gromov:2010km} and involves only two unknown functions which we denote $Q_1$ and $Q_2$:

\begin{align}
T_{1,s}=C\left|\begin{array}{cc}
\label{eq:newQ}
Q_1^{[s]} & \bar Q_1^{[-s]} \\
Q_2^{[s]} & \bar Q_2^{[-s]} \\
\end{array}\right|\;.
\end{align}
In this way we are able to replace the infinite set of $Y_m$ functions ($m=2,3,\dots$) by two functions $Q_1(u)$ and $Q_2(u)$.
Now the problem is reduced to finding an ansatz for the functions $Q_1,Q_2$ entering \eqref{eq:newQ}. The main requirement for this ansatz is that the $Y_{1,m}$ generated by \eq{YthroughTfull}, \eq{eq:newQ} should have the correct asymptotics at large real $u$ given by \eqref{bbYas}.
For small angles the asymptotics is $\frac{1}{m^2-1}$ and the corresponding ansatz for the Q-functions is known \cite{Gromov:2012eu}.
Here we present an ansatz which works also in a deformed case with nontrivial twists.

 The ansatz also has to ensure the correct analytical properties of the Y-functions which are dictated by the integral equations \eq{eq:TBAfinaleq3}. First of all, the $Y_{1,m}$ functions should be analytic inside the strip $|\text{Im}\ u|<\frac{m-1}{2}$ and even as functions of $u$. The term with $\delta_{m,2}$ in \eq{eq:TBAfinaleq3} can be reproduced if $Y_{1,2}(u)$ has branch cuts starting at $u=i/2\pm 2g$ and $u=-i/2\pm 2g$.

 Our proposal for $Q$-functions meeting these requirements is:
\begin{align}
\label{Qans1}
&Q_1=\bar Q_1=e^{+\theta (u-i \cG(u))},\\
\label{Rans1}
&Q_2=\bar Q_2=e^{-\theta (u-i \cG(u))},
\end{align}
where $\cG(u)$ should be a function with a branch cut on the real axis in order to satisfy the properties of T-functions listed above.
Note that the asymptotics \eqref{bbYas} of $Y$-functions
is automatically satisfied for any $\cG(u)$ decaying at infinity. Finally, as $T_{1,s}$ are even and real functions (to ensure the same properties for Y-functions), $\cG(u)$ should be odd and imaginary.

With this choice of $Q_1$ and $Q_2$ we can calculate $T_{1,s}$ from \eqref{eq:newQ}
where for consistency with \cite{Gromov:2012eu} in the small angle limit we choose $C=\frac{1}{2i\sin\theta}$
\beq
T_{1,s}=\frac{\sin(s-\cG^{[s]}+\cG^{[-s]})\theta}{\sin\theta}.
\label{eq:Texplicit}
\eeq

Discontinuity of the function $\cG$ can be found from the equation analogous to \eq{Y1m} for $m=2$ \cite{Gromov:2011cx}.
It reads
\beq
\label{eq:PhioverPsi}
\frac{ T_{1,1}^{+_+} T_{1,1}^{-_-}}{ T_{1,1}^{+_-} T_{1,1}^{-_+} }=r,\;\;\text{where}\;\;
r=\frac{1+1/Y_{2,2}}{1+Y_{1,1}}
\eeq
and
we denoted
\beq
T^{+_{\pm}}(u)=T(u+i/2\,{\pm\, i0})\qquad\text{and}\qquad T^{-_{\pm}}(u)=T(u-i/2\,{\pm\, i0})\;.
\label{eq:Tpmnotation}
\eeq
More explicitly, using the formula \eqref{eq:Texplicit} for $T_{1,1}$ one can write
\beq
r=\frac{\sin{\(1-\cG^{[+2]}+\slashed{\cG}-\rho/2 \)\theta}\sin{\(1+\cG^{[-2]}-\slashed{\cG}-\rho/2 \)\theta}}{\sin{\(1-\cG^{[+2]}+\slashed{\cG}+\rho/2 \)\theta}\sin{\(1+\cG^{[-2]}-\slashed{\cG}+\rho/2 \)\theta}},
\eeq
where $\slashed{\cG}(u)$ is the average of $\cG$ on both sides of the cut if $u$ is on the cut, and it is equal to $\cG(u)+\rho(u)/2$ away from the cut.
This allows to deduce the discontinuity of the function ${\cal G}$ with one real Zhukovsky cut
in terms of a combination \eq{eq:PhioverPsi} of ``fermionic" Y-functions $Y_{1,1}$ and $Y_{2,2}$.

Finally, for small $\theta$ the combinations $Q_1 \pm Q_2$ obtained from our ansatz nicely match\footnote{As $T_{1,s}$ are given by a determinant, we are free to replace $Q_{1,2}$ by their linear combinations} (up to overall factors) the Q-functions in the small angles case \cite{Gromov:2012eu}, where $Q_1=1$ and $Q_2=-iu-\cG(u)$.
\subsection{Expansion in the near-BPS case}
The ansatz presented in the previous subsection is valid for a general, not necessarily near-BPS situation. Here we will apply it to the case of $\phi\approx\theta$ (i.e. small $\epsilon$) studied in this paper.

As we have seen above, the solution for $Y$-functions is completely defined by a single function $\cG(u)$, which we will call the resolvent. For the goals of this paper we only need to know $\cG$ up to the linear in $\epsilon$ terms inclusive. Our proposal for the resolvent is
\beq
	\cG(u)=\frac{1}{2\pi i}\int\limits^{2g}_{-2g}dv\frac{\rho(v)}{u-v}+\epsilon\sum\limits_{a\ne 0}\frac{b_a}{u-i a/2}\;.
	\label{Gwpoles}
\eeq
The first term creates a short branch cut\footnote{i.e. a cut from $-2g$ to $2g$.} in $\cG(u)$, which translates into the branch cuts of $\cY_m$. The discontinuity of the resolvent across this cut is the density $\rho$:
\beq
\rho(u)=G(u-i0)-G(u+i0).
\eeq
The second term in \eqref{Gwpoles} produces poles at $\pm i/2$  with residues proportional to $\epsilon$ in Y-functions, which account for the term $\epsilon \pi \mathbf{s} \mathbb{C}_m$ in \eqref{eq:TBAfinaleq3}.

 One can see that the properties of $T_{1,m}$ being real and even imposes the following constraints on the density and poles: $\rho$ should be even and real as a function with a long cut, while $b_a=b_{-a}$ and $b_a=-b_{a}^*$.

Most of the equations in the paper are already expanded in $\epsilon$, so it is convenient to introduce expanded to the leading order versions of the quantities above.
The leading order part of the resolvent is\footnote{The density $\rho$ contains both the leading order in $\epsilon$ part and the linear correction, however, in this paper we will never need to deal with this correction. Hence, we will denote the full density and its leading order part by the same letter $\rho$ hoping that this will not cause any confusion.}
\beq
	G(u)=\frac{1}{2\pi i}\int\limits^{2g}_{-2g}dv\frac{\rho(v)}{u-v}.
	\label{eq:Gdefinition}
\eeq

We also introduce the leading order $T$-functions $\cT_m$ related to the leading order $Y$-functions as
\beq
{\cY}_{m}=\frac{{\cT}_{m}^+ {\cT}_{m}^-}{{\cT}_{m+1} {\cT}_{m-1}}-1.
\label{eq:YthroughT0}
\eeq
Explicitly, the leading order part of \eqref{eq:Texplicit} gives
\begin{align}
\cT_{s}=\frac{\sin{(s-G^{[s]}+G^{[-s]})\phi_0}}{\sin\phi_0}.
\label{eq:Texplicit0}
\end{align}

\subsection{Final reduction to FiNLIE}

We now use the ansatz discussed above and finalize the reduction of the initial Bremsstrahlung TBA system
to a finite set of equations. The remaining steps in the derivation are analogous to \cite{Gromov:2012eu} so we will be brief here (more details are given in Appendix \ref{sec:DerivationofFiNLIE}).

The first two TBA equations \eqref{eq:TBAfinaleq1}, \eqref{eq:TBAfinaleq2} contain the ``fermionic'' functions $\Phi$ and $\Psi$ in the left hand side. In order to deal with them, notice that after plugging the expansion \eqref{yexp1} of Y-functions into $r$ defined by \eqref{eq:PhioverPsi} one gets $r=\Phi/\Psi$. Thus the equation \eqref{eq:PhioverPsi} at the leading order becomes

\beq
\frac{\Phi}{\Psi}=\frac{{\cal T}_1^{+_+} {\cal T}_1^{-_-}}{ {\cal T}_1^{+_-} {\cal T}_1^{-_+} },
\label{eq:PhioverPsi1}
\eeq
 where the notation analogous to \eqref{eq:Tpmnotation} is used.
The equation \eqref{eq:PhioverPsi1} allows us to introduce another quantity which will play an important role in the FiNLIE:
\begin{align}
\eta\equiv \frac{\Psi {\cal T}_2}{{\cal T}_1^{-_+}{\cal T}_1^{+_-}}=\frac{\Phi {\cal T}_2}{{\cal T}_1^{-_-}{\cal T}_1^{+_+}}.
\label{eq:etadef}
\end{align}

Using this definition and the explicit form of $\cT_m$ \eqref{eq:Texplicit0} we are able to express $\Psi$ and $\Phi$ in terms of $\eta$, $\rho$ and $G$. Then we plug them into the first two TBA equations and get the first two FiNLIE equations \eq{eq:FiNLIE1}, \eq{eq:FiNLIE2} which are given below.

To get the third FiNLIE equation we plug the explicit form of the $\cY_m$ functions expressed through $\cT_m$ using \eqref{eq:YthroughT0} and \eqref{eq:Texplicit} into the equation for $\Delta_a$ (Eq. \eq{eq:TBAfinaleq5}). This equation then greatly simplifies (for a detailed derivation see Appendix \ref{sec:DerivationofFiNLIE}) and we find
\begin{align}
\Delta_a=\tilde K_a \hat *\log\eta+\left.\log\frac{{\cal T}_a}{\sin a\theta\cot\theta}\right|_{u=0}\;.
\label{eq:etaeq}
\end{align}
Combining this with the last equation of Bremsstrahlung TBA \eqref{eq:TBAfinaleq6} we obtain \eqref{eq:FiNLIE3}.

In summary, the FiNLIE equations read:
 \newline

 \fbox{
  \addtolength{\linewidth}{-2\fboxsep}%
  \addtolength{\linewidth}{-2\fboxrule}%
 \begin{minipage}{\linewidth}
\begin{align}
&{\eta}\frac{\sin{\theta\rho}}{\sin\theta}=- \sum\limits_a\pi \mathbb{C}_a\hat K_{a},&
\label{eq:FiNLIE1}\\
&\eta\frac{\cos{\theta\rho}\cos{(2-G^++G^-)\theta}-\cos{(2\slashed{G}-G^+-G^-)\theta}}{\sin\theta\sin{(2-G^++G^-)\theta}}=&
\nonumber
\\
\label{eq:FiNLIE2}
&= \mathbf{s}*\left[-2\frac{ {\cal X}_2 }{1+{\cal Y}_2 }+\pi(\hat K_a^+-\hat K_a^-)\mathbb{C}_a-\pi \delta(u) \mathbb{C}_1\right],&\\
&\mathbb{C}_a=(-1)^a a {\cal T}_a(0)\left(\sqrt{1+\frac{a^2}{16g^2}}-\frac{a}{4g}\right)^{2+2L} \exp\left[\tilde K_a \hat *\log\left(\eta\frac{\sinh{2\pi u}}{2\pi u}\right)\right].
\label{eq:FiNLIE3}
\end{align}
\end{minipage}
}
\newline

Here $\slashed{G}(u)$ is the average of the resolvent on both sides of the cut if $u$ is on the cut, and it is equal to $G(u)+\rho(u)/2$ away from the cut. Other notation and the kernels can be found in the Appendix \ref{sec:notationandconventions}.

Our FiNLIE is a set of equations for functions $\rho(u), \ \eta(u)$ and the coefficients $\mathbb{C}_a$ (we remind that $G$ is obtained from $\rho$ according to \eq{eq:Gdefinition}). As written this is a closed system of equations up to one subtlety. Namely, the r.h.s. of the second equation, \eq{eq:FiNLIE2}, also includes an unknown function ${\cal X}_2$ which should contain the linear in $\epsilon$ correction to $\rho$. This correction obeys an equation which is straightforward to derive by the same methods as in \cite{Gromov:2012eu}\footnote{see eq. (\textbf{F3}) in section 3.5 of \cite{Gromov:2012eu}}. However, in fact we will not need this equation in the following, so we do not write it.
It is replaced by a certain simple analyticity condition described in the next section. This condition is a simple
consequence of the novel ${\bf P}\mu$-system formulation in Sec. \ref{sec:Pmu}, which is, however, very hard to prove starting directly from TBA.

Finally, the FiNLIE should be also supplemented by a relation which determines the residues of the resolvent at $u=ia/2$, i.e. the coefficients $b_a$ which we introduced in \eq{Gwpoles}. To derive it we compare residues at $ia/2$ of both sides of the third equation \eq{eq:TBAfinaleq3} in the Bremsstrahlung TBA system. This gives a recursion relation of the form
\beq
q_a b_{a-2}-(q_a+p_a)b_a+p_a b_{a+2}= \mathbb{C}_a,
\label{eq:recrelation}
\eeq
where $q_a$ and $p_a$ depend on the values of the resolvent at the points $ia/2$, $i(a\pm 2)/2$, and are defined in the Appendix \ref{sec:fixba} in which
the derivation of \eq{eq:recrelation} is discussed.

In the next section we will construct an analytic solution of this FiNLIE, leading to an explicit expression for the energy.


\section{Solving the FiNLIE: analytical ansatz}
\label{sec:Solving}
In the previous sections we presented the FiNLIE - a system of equations for $\mathbb{C}_a,\rho,\eta$. Following the spirit of \cite{Gromov:2012eu}, in order to solve it we will analyse the analytical properties of $\eta$ and $\rho$ as functions in the whole complex plane. We will parametrize these functions in terms of auxiliary Bethe roots, for which we will obtain a set of Bethe equations. Then we solve them using Baxter equation techniques and obtain the result for the anomalous dimension $\Gamma_L(g)$.

\subsection{Analytical ansatz for $\eta$ and $\rho$}

In this section we will explore the analytical properties of $\rho$ and $\eta$. Although
one would prefer to derive them starting from the FiNLIE, there seems to be no easy way to do this.
 Instead we make a conjecture that the key quantities entering the FiNLIE do not have infinitely many Zhukovsky cuts and then
  justify it using the novel ${\bf P}\mu$-system  \cite{PmuPRL} formulation
  in section \ref{sec:Pmu}. We will show that this statement follows almost trivially
  from the ${\bf P}\mu$-system techniques. In this way we also justify similar assumptions made in \cite{Gromov:2012eu} without a proof.

The main assumption is that $\eta$ has simple poles at $ia/2$ for $a\in \mathbb{Z}\setminus\{0\}$, and  $\eta^2(u)$ is a meromorphic function in the whole complex plane.
Then, taking into account that $\eta$ is even we can write the following representation
\beq
\eta^2(u)=(\cos\theta)^2\prod
\limits_{k\ne 0}
\frac{u^2-u_{k}^2}{u^2+k^2/4},
\label{eq:etau}
\eeq
where the product goes from $-\infty$ to $\infty$. The prefactor $(\cos\theta)^2$ comes from the asymptotics
\beq
\eta(u)\rightarrow\cos\theta, \ u\rightarrow \infty,
\eeq
which is easily seen from the definition \eqref{eq:etadef}.

In \cite{Gromov:2012eu}, where the case $\theta=0$ was considered,
 $\eta$ was a meromorphic function with poles at $ia/2$ for nonzero integer $a$. In our case $\eta^2$ rather than $\eta$ is meromorphic.
 The analyticity of $\eta$ in the $\theta=0$ limit is recovered as pairs of zeros $u_k$ collide
 and produce a double zero. We enumerate these zeros $u_k$ in
  such a way that the colliding pairs are $u_k$ and $-u_{-k}$. For large $k$ the value of the
   factors under the product in \eqref{eq:etau} should approach $1$,
 meaning that the roots accumulate close to the half-integer points on the imaginary axis:
\begin{align}
&u_{k} \rightarrow ik/2,\\ \nn
-&u_{-k} \rightarrow ik/2, \ k\rightarrow+\infty.
\end{align}
Thus $\eta$ has an infinite number of square-root cuts, each going between
$u_{k}$ and $-u_{-k}$, located close to $ik/2$. We will refer to these cuts as $S$-cuts, and they are shown in Fig. \ref{pic:balalaika}.

\FIGURE[ht]
{\label{pic:balalaika}

    \begin{tabular}{cc}
    \includegraphics[scale=0.9]{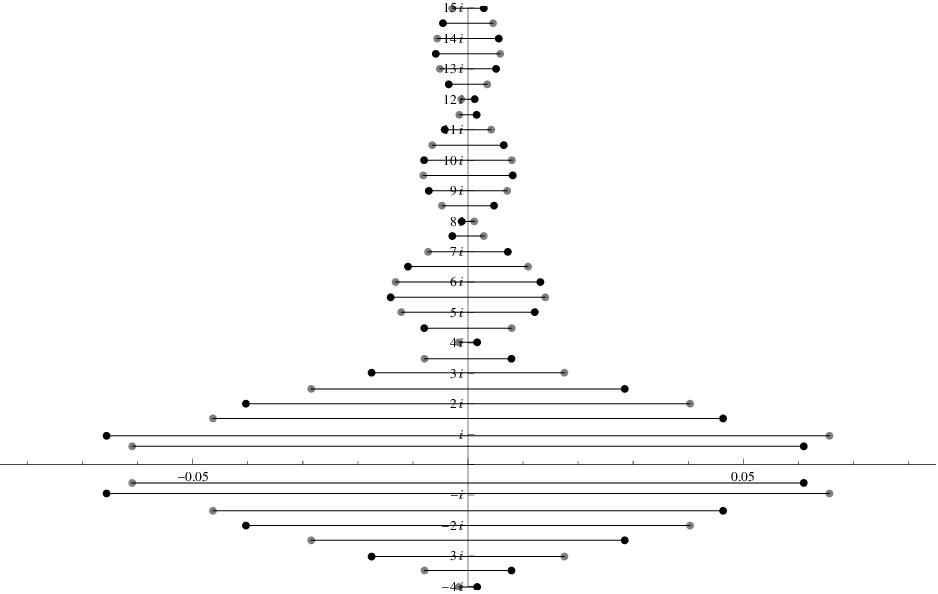}\\
    \end{tabular}
    \caption{
    \textbf{The S-cuts of $\eta(u)$ and $\rho(u)$.} The function $\eta(u)$ has an infinite number of square root cuts, which we call S-cuts, each connecting $u_k$ (black dots) to $-u_{-k}$ (grey dots). For $\rho(u)$ these cuts are logarithmic. The full set of branch points consists of $\{u_k\}$ together with $\{-u_k\}$.
    }
    }

Now let us explore the properties of the density $\rho$. Every kernel $\hat K_a$ in the right hand side of the first FiNLIE equation
\eqref{eq:FiNLIE1} is proportional to $\sqrt{u^2-4g^2}$,
so the whole expression has a cut from $-2g$ to $2g$, which we will call the Z-cut.
First, let us note that $\rho$ is defined as a discontinuity of the resolvent $G$
and as such it simply changes its sign when passing the $Z$-cut and so the Z-cut is already taken
care of by the ${\sin{\theta\rho}}$ multiplier.

It is left to understand the behavior of $\rho$ when we go through an S-cut.
As the combination $\eta \sin{\theta\rho}$
has no S-cuts due to \eqref{eq:FiNLIE1}
and since $\eta$ does have infinitely many S-cuts, it must be
that $\sin\theta\rho$ changes its sign simultaneously
with $\eta$ when we go through any S-cut leaving the whole expression unchanged.
Next, let us show that $\cos{\theta\rho}$ also changes sign on an $S$-cut.
Indeed, in the second FiNLIE equation
\eq{eq:FiNLIE2}
 the right hand side does not have any
$S$-cuts and the left hand side can be expanded into a sum of
terms proportional to $\eta\sin{\theta\rho}$ and $\eta\cos{\theta\rho}$.
Since from the first FiNLIE equation \eq{eq:FiNLIE1}
we know that $\eta\sin\theta\rho$ does not branch on an S-cut,
the same should be true for $\eta\cos{\theta\rho}$.
Again, since $\eta$ changes its sign on $S$-cuts,
the same should hold for  $\cos{\theta\rho}$.

This means that on the $Z$-cut $\rho$ changes its sign and on an $S$-cut it is shifted as
 \beq
 \rho\rightarrow \rho+\pi/\theta.
 \label{eq:rhoshift}
 \eeq
The transformation properties of different quantities with respect to transitions through $Z$- and $S$-cuts can be summarized into the following table:

\begin{center}
\begin{tabular}{|l|c|c|c|r|}
  \hline   & $\eta$  & $\rho$ & $\cos{\theta\rho}$ & $\sin{\theta\rho}$ \\
  \hline $S$-cut  &  $-\eta$  & $\rho+\pi/\theta$ & $-\cos{\theta\rho}$ & $-\sin{\theta\rho}$ \\
  \hline $Z$-cut &  $\eta$ & $-\rho$ & $\cos{\theta\rho}$ & $-\sin{\theta\rho}$
   \\
  \hline
  \end{tabular}
  \end{center}

Having understood the transformation properties of $\rho$ on both types of cuts, let us try and build out of $\rho$ a quantity which would be meromorphic. First of all, to this end it is convenient to consider $\rho$ as a function of Zhukovsky transformed variable $x(u)$ such that
\beq
	\frac{u}{g}=x(u)+\ofrac{x(u)}.
\eeq
It is easy to see that Zhukovsky transformation resolves the $Z$-cut: two sheets of the Riemann surface connected by the cut in variable $u$ become the interior and the exterior of the unit circle in variable $x$.
Thus as a function of $x$ the density has only $S$-cuts. Moreover, since on an $S$-cut $\rho$ transforms to $\rho+\pi/\theta$, the combination $e^{2i\theta\rho(x)}$ is meromorphic in $\mathbb{C}\setminus \{0\}$.
Going under the $Z$-cut in variable $u$ is equivalent to $x\rightarrow 1/x$ transformation in variable $x$, hence the property of $\rho$ changing sign on the $Z$-cut now reads

\beq
e^{2i\theta\rho(x)}=1/e^{2i\theta\rho(1/x)}.
\label{eq:exprho}
\eeq

Being meromorphic, $e^{2i\theta\rho(x)}$ is completely characterized by its zeros and poles (and asymptotics).
Let us call the zeros outside the unit circle $x_{k,+}$ and the zeros inside it $1/x_{k,-}$. Then from \eqref{eq:exprho} one can see that the poles of $e^{2i\theta\rho(x)}$ are $1/x_{k,+}$ and $x_{k,-}$. These poles and zeros are shown in Fig. \ref{pic:zeros}.

\FIGURE[ht]
{\label{pic:zeros}
\begin{tabular}{cc}
    \includegraphics[scale=0.2]{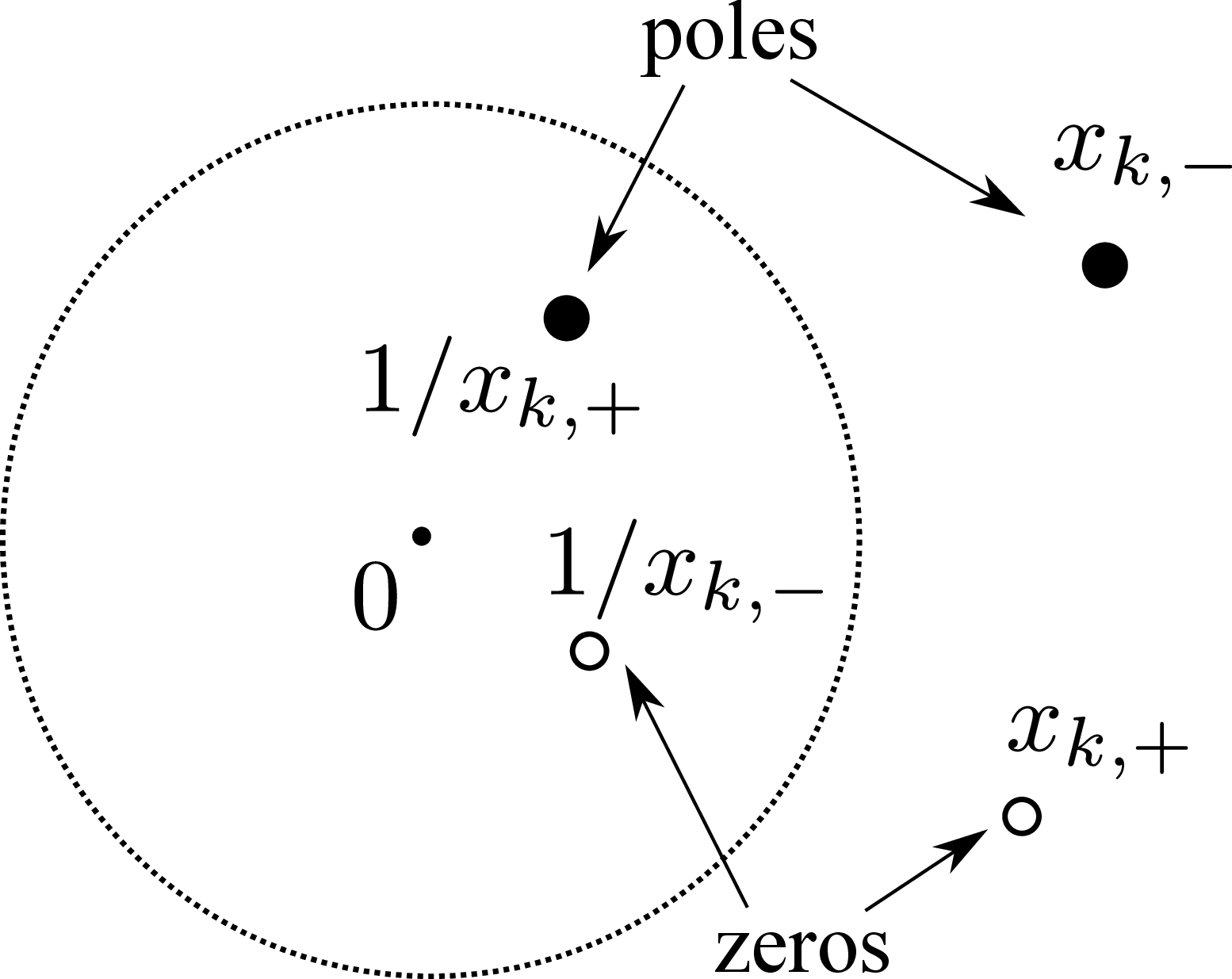}\\
    \end{tabular}
    \caption{
    \textbf{The singularities of $\rho$.}
Poles and zeros of $e^{2i\theta\rho(x)}$ inside and outside the unit circle. The density $\rho$ has logarithmic singularities at these points, which are in fact images of $\pm u_k$ under the Zhukovsky map.
    }
    }

It is convenient to introduce bookkeeping functions which encode $x_{k,\pm}$

\beq
\bQ_{\pm}(x)=\prod_{k\ne 0} \frac{x_{k,\pm}-x}{x_{k,\pm}}\;.
\label{eq:Qdef}
\eeq
We denote their analytical continuation under the $Z$-cut by adding a tilde, i.e. $\tilde \bQ_{\pm}(x)= \bQ_{\pm}(1/x)$.
 Knowing the zeros and the poles of $e^{2i\theta\rho(x)}$ and taking into account that $\rho\rightarrow 0$ as $x\rightarrow\infty$, we can reconstruct it uniquely as

\beq
e^{2i\theta\rho(x)}=\frac{\bQ_+\tilde \bQ_-}{\bQ_-\tilde \bQ_+}.
\label{eq:rhoansatz2}
\eeq

Using this representation for $\rho$ we can fix $\eta$ completely. Indeed, as discussed above the l.h.s. of the first FiNLIE equation \eqref{eq:FiNLIE1} does not have $S$-cuts. On the other hand, we can use \eqref{eq:rhoansatz2} to write it in terms of ${\bf Q}$ as
\beq
\eta\sin\theta\rho=\frac{\eta}{2i}\frac{\tilde \bQ_- \bQ_+-\tilde \bQ_+\bQ_-}{\sqrt{\bQ_-\tilde \bQ_- \bQ_+\tilde \bQ_+}}.
\label{eq:sinrepr}
\eeq
Thus the square root in the denominator of \eqref{eq:sinrepr} should completely cancel the numerator of $\eta$ which is equal to $\prod\limits_{k\neq 0}\sqrt{u^2-u_{k}^2}$.
The asymptotics of the numerator should be $\sinh{2\pi u}$, because $\eta(u)$ is finite at infinity. Thus the zeros of $\sqrt{\bQ_-\tilde \bQ_- \bQ_+\tilde \bQ_+}$ should approach the zeros of $\sinh{2\pi u}$ and there is a way to enumerate $x_{k,+}$ so that $x_{k,+}\rightarrow ik/2$ at large $k$. Since $\rho$ is odd\footnote{as a function with a short $Z$-cut}, the zeros of the numerator and the denominator of \eqref{eq:rhoansatz2} should map onto each other as sets under $x\rightarrow -x$. In particular, considering only the zeros outside the unit circle, the zeros of $\bQ_{+}$ map onto zeros of $\bQ_{-}$. Then it is possible to enumerate the zeros of $\bQ_-$ (i.e.$\;x_{k,-}$) so that $x_{k,+}=-x_{-k,-}$. Notice that now $u(x_{k,+}),\;u(x_{-k,-})\rightarrow ik/2$ as $k\rightarrow\infty$. Introducing $v_k$ such that $x(v_k)=x_{k,+}$, we can write
\beq
\bQ_{+}\bQ_{-}\tilde\bQ_{+}\tilde\bQ_{-}=\prod\limits_{k\ne 0}\frac{x-x_{k,+}}{x_{k,+}}\frac{x-x_{-k,-}}{x_{-k,-}}\(1-\frac{1}{x x_{k,+}}\) \(1-\frac{1}{x x_{-k,-}}\)=\prod\limits_{k\ne 0}\frac{u^2-v_k^2}{-g^2x_{k,+}^2}
\eeq
Comparing this product with the product in the numerator of $\eta^2$ we see that $u_k$ and $v_k$ coincide as sets. Up to relabelling we can set $v_k=u_k$, thus establishing the relation
\beq
u_k/g=x_{k,+}+1/x_{k,+}=-x_{-k,-}-1/x_{-k,-}.
\eeq

Finally, we notice that the formula \eqref{eq:rhoansatz2} allows us to find the resolvent in terms of $\bQ_{\pm}$ without performing the integration which is prescribed by \eqref{eq:Gdefinition}. Indeed, the function $G(u)$ defined as $e^{i\theta G(u)}=\sqrt{\tilde\bQ_+\(x(u)\) /\tilde\bQ_-\(x(u)\)}$ decays at infinity, doesn't have poles
on the main sheet ($|x|>1$) and has a $Z$-cut with the discontinuity $\rho$, the same as the resolvent. Hence by Liouville's theorem it coincides with the resolvent.

Let us summarize the results of this section:
\\
\\
 \fbox{
  \addtolength{\linewidth}{-2\fboxsep}%
  \addtolength{\linewidth}{-2\fboxrule}%
 \begin{minipage}{\linewidth}
\begin{align}
e^{i\theta\rho}=\sqrt{\frac{\bQ_+\tilde \bQ_-}{\bQ_-\tilde \bQ_+}},&\;\;\;\;
\eta=\cos\theta\frac{\sqrt{\bQ_+\bQ_- \tilde\bQ_+ \tilde\bQ_-}}{\tilde C \frac{\sinh 2\pi u}{2\pi u}}\;,
\label{eq:etansatz}
\\
 e^{i\theta G}&=\sqrt{\tilde\bQ_+ /\tilde\bQ_-}\;.
 \label{eq:GthroughQ}
\end{align}
\end{minipage}
}
\\
\\
\\
In order to rewrite the ansatz for $\eta$ \eqref{eq:etau} in the form above we used the identity
\beq
\frac{\sinh{2\pi u}}{2\pi u}=\prod\limits_{k=1}^\infty\frac{u^2+k^2/4}{k^2/4}
\eeq
and we also introduced
\beq
\tilde C=\prod\limits_{k=1}^\infty\frac{-k^2/4}{g^2 x_{k,+} x_{k,-}}.
\eeq
We managed to write all the key quantities in terms of an infinite number of roots $u_{k}$.
By plugging these expressions into the FiNLIE equations in the next couple of sections
we will find a closed set of Bethe-like equations for these roots.

\subsection{Fixing residues of $\eta$}
\label{sec:relation}

Here we will find a relation for the residues of $\eta$
and as a result
establish an important relation between values of $\rho$ and $G$ at half-integer points on the imaginary axis which will be used in the next section to derive an auxiliary Bethe-like equation.
Here we only outline the main steps, with more details given in Appendix \ref{sec:FixingApp} \footnote{the calculation is analogous to that done in sections 3.2.1 and 4.2 of \cite{Gromov:2012eu}}.

First, we will use the second FiNLIE equation \eq{eq:FiNLIE2}, i.e.
\beqa
&&\eta\frac{\cos{\theta\rho}\cos{(2-G^++G^-)\theta}-\cos{(2\slashed{G}-G^+-G^-)\theta}}{\sin\theta\sin{(2-G^++G^-)\theta}}=\\ \nn &&
=\mathbf{s}*\left[-2\frac{ {\cal X}_2 }{1+{\cal Y}_2 }+\pi(\hat K_a^+-\hat K_a^-)\mathbb{C}_a-\pi \delta(u) \mathbb{C}_1\right]\;,
\eeqa
to compute the residue of $\eta \cos\theta\rho$ at $ia/2$. According to our assumptions about analytical properties of $\eta$, both sides of this equation have poles at $ia/2$.
Using the identity \eqref{eq:invtranslationdeformed} we can get rid of the convolution with ${\bf s}$. Then the residues at the pole on both sides can be expressed through $\mathbb{C}_a$, $G(ia/2)$ and the residues of $\eta\cos\theta\rho$ and ${\cal X}_a$.
 Due to the presence of ${\cal X}_2$, the residue of the r.h.s. appears to depend on $b_a$ (see the definition of $G$, where the poles with residues $b_a$ appear in the first order in $\epsilon$). However, as in \cite{Gromov:2012eu}, the dependence on $b_a$ can be completely eliminated by taking into account the recursion relation \eqref{eq:recrelation}. Thus we can regard the second FiNLIE equation \eq{eq:FiNLIE2} as an equation for the residues of $\eta\cos{\theta\rho}$ which produces as shown in Appendix \ref{sec:FixingApp}:
\beq
\underset{u=i a/2}{\text{Res }}(\eta\cos{\theta\rho})=\frac{1}{2i}\mathbb{C}_a \frac{\sin\theta}{\tan{(2 G(ia/2)\theta-a\theta)}}.
\label{etacos}
\eeq

In addition, let us make use of the first FiNLIE equation \eq{eq:FiNLIE1},
\beq
	{\eta}\frac{\sin{\theta\rho}}{\sin\theta}=- \sum\limits_a\pi \mathbb{C}_a\hat K_{a}.
\eeq
Equating the residues of the poles at $ia/2$ on both sides gives us at once
\beq
\underset{u=i a/2}{\text{Res }}(\eta\sin{\theta\rho})=-\frac{1}{2i}  \mathbb{C}_a \sin\theta.
\label{etasin}
\eeq

The equations \eq{etacos} and \eq{etasin} that we have just derived allow us to relate $\rho$ and $G$ at $u=ia/2$. Since we assume that $\rho$ is regular at $ia/2$ and the pole comes from $\eta$, from these two equations it is easy to see that
\beq
\tan\theta\rho(ia/2)=\tan{(a\theta-2 G(ia/2)\theta)},
\eeq
leading to
\beq
	\theta\rho(ia/2)={a\theta-2 \theta G(ia/2)}+ \pi n,\ \ \ n\in\mathbb{Z}\ .
\label{eq:densityequation}
\eeq
We can also write this equation as\footnote{For odd $n$ in \eq{eq:densityequation} we would get minus in r.h.s. of \eq{eq:exprhoeq}, but this is incompatible with the small $\theta$ limit (see Eq. (103) in \cite{Gromov:2012eu}).}
 \beq
 \exp\left[i\theta\rho(ia/2)+2i\theta G(ia/2)-ia\theta\right]=1\;.
 \label{eq:exprhoeq}
 \eeq
This relation already constrains the set of our parameters $u_k$ and will be very useful in the next section.

\subsection{Effective Bethe equations}
Above we have parametrized $\rho$ and $\eta$ in terms of two families of roots $x_{k,\pm}$. Here we will show that these roots satisfy a set of Bethe-like equations, which will be solved in the subsequent sections.
In \cite{Gromov:2012eu} the effective Bethe equation was derived by substituting the ansatz for $\eta$ into the FiNLIE equation for $\bbC_a$. In our case this is Eq. \eq{eq:FiNLIE3} which reads
\beq
	\mathbb{C}_a=(-1)^a a {\cal T}_a(0)\left(\sqrt{1+\frac{a^2}{16g^2}}-\frac{a}{4g}\right)^{2+2L} \exp\left[\tilde K_a \hat *\log\left(\eta\frac{\sinh{2\pi u}}{2\pi u}\right)\right].
\eeq
Plugging into this equation our ansatz for $\eta$ \eqref{eq:etansatz} and following the same steps as in \cite{Gromov:2012eu}, we get
\beq
1=\left(\frac{i}{y_a}\right)^{2L+2}\sqrt{\frac{\tilde \bQ_+\tilde \bQ_-}{\bQ_+ \bQ_-}}\;\;,\;\;\text{with }y_a\equiv x(ia/2).
\label{eq:Bethe22}
\eeq
In the $\theta=0$ limit $\bQ_+=\bQ_-$ and this equation coincides with the effective Bethe equation in \cite{Gromov:2012eu}.
In addition to \eq{eq:exprhoeq} this equation allows to fix completely
 all the roots $x_k$, thus providing the full solution to the problem.

Indeed, expressing $\rho$ and $G$ in \eqref{eq:exprhoeq} through $\bQ_{\pm}$ by means of \eqref{eq:etansatz} and \eqref{eq:GthroughQ} we obtain
\beq
1=e^{-2 u\theta}\sqrt{\frac{\bQ_+\tilde \bQ_+}{\bQ_-\tilde \bQ_-}}\;\;,\;\;u=ia/2.
\label{eq:Bethe11}
\eeq
Notice that this equation contains $\theta$ (as opposed to \eqref{eq:Bethe22}) and tells us how the two families of roots are separated. In the $\theta=0$ limit it has a trivial solution $\bQ_+=\bQ_-$, causing the roots $x_{k,+}$ and $-x_{-k,-}$ to collide and producing double zeros in $\eta^2$, thus making $\eta$ meromorphic.

Multiplying \eqref{eq:Bethe11} and \eqref{eq:Bethe22} we get rid of the square root and finally obtain the following auxiliary Bethe equations:
\begin{align}
\label{eq:Bethesimpl1}
\boxed{
1=e^{-ia\theta}\left(\frac{i}{y_a}\right)^{2L+2}\frac{\tilde \bQ_+(y_a)}{\bQ_-(y_a)}.
}
\end{align}
We remind that $y_a$ stands for $x(ia/2),\;a\in{\mathbb Z}$.
From this equation we can find the Baxter polynomials $\bQ_\pm$. For that
in the next section we will use the Baxter equation.

\subsection{Baxter equations}

At this point everything we want to know about the system is parametrized in terms of two infinite series of roots $x_{k,\pm}$. These roots are governed by the effective Bethe equations \eqref{eq:Bethesimpl1}, and to solve them we will apply Baxter equation techniques, similarly to \cite{Gromov:2012eu}.

Namely, let us construct the function
\beq
\mathbf{T}(x)=e^{+2g\theta x}x^{L+1}\bQ_{-}(x)+(-1)^L\frac{e^{-2g\theta/x}}{x^{L+1}}\tilde\bQ_{+}(x).
\label{eq:baxterconsider}
\eeq
which encodes the whole set of auxiliary Bethe roots $x_k$. We will call $\bT(x)$ the Baxter function. Due to the Bethe equations \eqref{eq:Bethesimpl1} we have $\bT(y_a)=0$. In addition, the relation $\bQ_{\pm}(-x)=\bQ_{\mp}(x)$ means that $\bT$ has a symmetry
\beq\label{Tinv}
	\mathbf{T}(-1/x)=-\mathbf{T}(x)
\eeq

Let us now clarify the asymptotics of $\mathbf{T}(x)$. It is easy to see from the definitions \eqref{eq:Qdef} and \eqref{eq:etadef} that $\tilde{\bQ}_{\pm}\rightarrow 1$ and $\eta\rightarrow\cos\theta$ at large $x$. Moreover, since $\bQ_{\pm}(-x)=\bQ_{\mp}(x)$, the asymptotics of $\bQ_+$ and $\bQ_-$ at large $u$ are the same and from \eqref{eq:etansatz} we get
\beq
	\bQ_{\pm}\sim  \tilde C\frac{\sinh{2\pi u}}{2\pi u},\ u\rightarrow +\infty\;.
\eeq
Therefore the second term in \eqref{eq:baxterconsider} is suppressed\footnote{Strictly speaking this is so for $-\pi<\theta<\pi$. Using periodicity in $\theta$
we can always restrict ourselves to this range.} compared to the first one and the asymptotics of the whole expression at large $x$ is $\bT(x)\sim x^{L}e^{2g(\pi+\theta)x}$. Then from \eq{Tinv} we can find the asymptotics of $\bT(x)$ at $x\to0$, and combining
all these analytical properties together we can fix $\bT$ uniquely to be
\beqa
\mathbf{T}(x)=\sinh(2\pi u)e^{2g\theta(x-1/x)}P_L(x),
\label{eq:conjecture}
\eeqa
where $P_L(x)$ should be a rational function with behavior $\sim x^L$ at infinity. Since $\bT(x)$ should not have singularities apart from $x=0$ and $x=\infty$, the function $P_L$ must be a polynomial in $x$ and $1/x$. Moreover, \eq{Tinv} means that
$P_L(-1/x)=P_L(x)$
and hence we can write
\beq
P_L(x)=C_1 x^L+C_2 x^{L-1}\dots +(-1)^L C_1 x^{-L}.
\eeq
To find $\bT(x)$ explicitly it only remains to determine the coefficients $C_i$. This is straightforward to do by imposing the condition that the r.h.s. of \eqref{eq:conjecture} does not contain powers of $x$ from $-L$ to $L$ in its Laurent expansion (as follows from \eqref{eq:baxterconsider})
which must be the case since $\bQ_-$ is regular at the origin.

\subsection{The energy}
Before proceeding with fixing completely the Baxter function $\bT(x)$, let us explain how to extract from it the value of the energy. To do this we will use the first FINLIE equation \eq{eq:FiNLIE1}-- at large $u$ each of the kernels $\hat K_a$ in its r.h.s. decays as $1/u$, so the whole sum is proportional to the sum in the definition of $\Gamma_L(g)$ \eqref{eq:Gamma}. Hence we get
\beq
\Gamma_L(g)=-i(\phi-\theta)\theta\lim_{u\to\infty}u\rho(u).
\eeq
The density is defined through $\bQ_{\pm}$ in \eqref{eq:etansatz}, so we can find the asymptotics of $\bQ_{\pm}$ from \eqref{eq:conjecture}, plug them into \eqref{eq:etansatz} and finally get

\beq
\Gamma_L(g)=-2(\phi-\theta)g\[-\frac{C_2}{2C_1}+\frac{c}{2}+g\theta\],
\label{eq:BfromC}
\eeq
where $c$ is the leading expansion coefficient of $\bQ_{\pm}$:
\beq\label{Qc}
\bQ_\pm(x) \simeq 1\mp cx\;\;,\;\;x\to 0.
\eeq
Notice that the coefficients $C_1,C_2$ are also encoded in $\bQ_\pm$: from \eq{eq:baxterconsider}, \eq{eq:conjecture} we find
\beq
	\bQ_\pm(x)\simeq \sinh(2\pi u) \[\frac{C_1}{x}\pm\frac{2g\theta C_1}{x^2}\mp\frac{C_2}{x^2}+\dots\]\;\;,\;\;x\to\infty\;.
\eeq
Now we have all the necessary tools to obtain the energy explicitly.

\subsection{The $L=0$ case}
Let us first discuss the $L=0$ case, because it is technically simpler. The function $P_L(x)$ from \eq{eq:conjecture} is then just a constant,
\beq
P_L(x)=C_1.
\eeq
To fix it we need to know the expansion of \eqref{eq:conjecture} in powers of $x$. Using that the exponent of $x+1/x$ is a generating
function for the modified Bessel functions of the first kind,
$
	e^{2\pi g\left(x+1/x \right)}=\sum\limits_{n=-\infty}^\infty I_n(4\pi g)x^{n}
$,
we get the expansion
\beq
\sinh\left(2\pi g(x+1/x)\right)e^{2g\theta(x-1/x)}=\sum\limits_{n=-\infty}^{+\infty}I_n^{\theta}x^n,
\label{eq:besselexpansion}
\eeq
where $I_n^\theta$ are the ``deformed'' Bessel functions
\beq
I_n^\theta=\frac{1}{2}I_{n}\(4\pi g\sqrt{1-\frac{\theta^2}{\pi^2}}\)\[
\(\sqrt{\frac{\pi+\theta}{\pi-\theta}}\)^{n}-
(-1)^n\(\sqrt{\frac{\pi-\theta}{\pi+\theta}}\)^{n}
\].
\label{eq:In}
\eeq
Below we will omit the argument of $I_n$, always assuming it to be the same as in \eqref{eq:In}.

The expansion \eqref{eq:besselexpansion} allows us to write the Baxter function \eqref{eq:conjecture} as
\beqa
\mathbf{T}(x)=e^{+2g\theta x}x\bQ_{-}(x)+\frac{e^{-2g\theta/x}}{x}\bQ_{+}(1/x)=
\nn C_1\sum_{n=-\infty}^{+\infty}I_{n}^\theta x^n.
\eeqa
We can now find $\bQ_-$ as the regular part of the Laurent expansion of $\mathbf{T}$:
\beq
\label{QmL0}
\bQ_{-}(x)=C_1\frac{e^{-2g\theta x}}{x}\sum_{n=1}^{+\infty}I_{n}^\theta x^n.
\eeq
From \eqref{eq:Qdef} we see that $\bQ_\pm(0)=1$, so setting $x=0$ in the last equation we fix $C_1$ as
\beq
	C_1=\frac{\sqrt{\pi^2-\theta^2}}{\pi I_1}.
\eeq
Since $L=0$ we have $C_2=0$, while the coefficient $c$ in \eq{Qc} is read off from \eq{QmL0}:
\beqa
c&=&-2 g \theta +\frac{2 \theta  }{\sqrt{\pi ^2-\theta ^2}}\frac{I_2}{I_1}\;.
\eeqa
Then from \eqref{eq:BfromC} we get the energy
\beq
	\Gamma_L
	=-2(\phi-\theta) \frac{\theta g}{\sqrt{\pi^2-\theta^2}}
	\frac{I_2\(\tilde\lambda^{1/2}\)}{I_1\(\tilde\lambda^{1/2}\)},\;\tilde{\lambda}=(4\pi g)^2\(1-\tfrac{\theta^2}{\pi^2}\).
	\label{eq:GammaL0}
\eeq
Remarkably, this is precisely the localization result of \cite{Correa:2012at}! This is the first successful check of our construction.

\subsection{Non-zero L}

Let us now find the explicit expression for the energy at any $L$.

First we need to compute the coefficients $C_k$, using the equation \eqref{eq:conjecture}. From \eq{eq:baxterconsider} we see that the left hand side of \eqref{eq:conjecture} should not contain terms with powers of $x$ from $-L$ to $L$, and also the coefficient of the $x^{L+1}$ term should be 1. After we expand the right hand side according to \eqref{eq:besselexpansion} this condition generates $2L+1$ equations for $2L+1$ variables $C_k$:

\begin{align}
\begin{cases}
\sum\limits_{k=-L}^L I_{m-k}^\theta C_{k+L+1}=0,\ m=-L+1\dots L,
\\
\sum\limits_{k=-L}^L I_{m-k}^\theta C_{k+L+1}=1,\ m=L+1.
\end{cases}
\end{align}
This linear system can be formulated in matrix form:
\beq
({\cal M}_{2L})_{ik} C_{k+L+1}=\delta_{i,L+1},
\label{eq:linsystem}
\eeq
where
\beq
{\cal M}_{N}=\begin{pmatrix}
I_1^{\theta}& I_0^{\theta}& \cdots & I_{2-N}^{\theta}  &I_{1-N}^{\theta}\\
I_2^{\theta}& I_1^{\theta}& \cdots & I_{3-N}^{\theta} &I_{2-N}^{\theta}\\
\vdots      &  \vdots     &\ddots & \vdots            &\vdots           \\
I_{N}^{\theta}& I_{N-1}^{\theta}& \cdots & I_{1}^{\theta}  &I_{0}^{\theta}\\
I_{N+1}^{\theta}& I_{N}^{\theta}& \cdots & I_{2}^{\theta} &I_{1}^{\theta}
\label{eq:M}
\end{pmatrix}.
\eeq
By Cramer's rule we obtain the solution
\beq
C_k=\frac{\det {\cal M}_{2L}^{(2L+1,k)}}{\det{\cal M}_{2L}},
\label{eq:Ck}
\eeq
where $ {\cal M}^{(a,b)}_N$ is the matrix obtained from ${\cal M}_N$ by deleting $a^{\text{th}}$ row and $b^{\text{th}}$ column. Plugging these coefficients into $P_L(x)$ we can combine it into a determinant again:
\beq
P_L(x)=\frac{1}{\det {\cal M}_{2L}}\left|\begin{matrix}
I_1^{\theta}& I_0^{\theta}& \cdots & I_{2-2L}^{\theta}  &I_{1-2L}^{\theta}\\
I_2^{\theta}& I_1^{\theta}& \cdots & I_{3-2L}^{\theta} &I_{2-2L}^{\theta}\\
\vdots      &  \vdots     &\ddots & \vdots            &\vdots           \\
I_{2L}^{\theta}& I_{2L-1}^{\theta}& \cdots & I_{1}^{\theta}  &I_{0}^{\theta}\\
x^{-L}& x^{1-L}& \cdots & x^{L-1} &x^{L}\\
\end{matrix}\right|.
\label{eq:PLrepr}
\eeq

Notice that now from \eq{eq:conjecture} we have the Baxter function $\bT(x)$ in a fully explicit form. In particular, one can easily find the functions $\bQ_\pm$ encoding the Bethe roots. Namely, $\bQ_-$ is the regular part of the Laurent expansion of $\mathbf{T}(x)$,
\beq
\bQ_-(x)=x^{-L-1}e^{-2g\theta x}\left[\mathbf{T}(x)\right]_+\ ,
\label{eq:QfromT}
\eeq
while $\bQ_+(x)=\bQ_-(-x)$.

It remains to find $c$ -- the coefficient of expansion of $\bQ_{\pm}$ which enters the expression for $\Gamma_L(g)$. Consider expansion of \eqref{eq:conjecture} around $x=0$, taking into account the definition of $\mathbf{T}$ \eqref{eq:baxterconsider}:
\beq
(1+2g\theta x+\dots)x^{L+1}(1+c x+\dots)+\text{negative powers}=\sum\limits_{n=-\infty}^{+\infty}I_n^{\theta}x^n \sum\limits_{k=-L}^L C_{k+L+1} x^k
\eeq
Equating the coefficients of $x^L$ on both sides we get
\beq
2g\theta+c=\sum\limits_{k=-L}^LI_{L+2-k}C_{k+L+1}\;.
\eeq
Plugging the solution for $C_k$ into the right hand side of the last equation we see that it combines nicely into a ratio of two determinants, resulting in
\beq
c=-2g\theta+\frac{\det {\cal M}_{2L+1}^{(2L+1,2L+2)}}{\det {\cal M}_{2L}}.
\label{eq:c}
\eeq
The determinants $\det {\cal M}_N^{(a,b)}$ satisfy a number of useful identities which we describe in Appendix \ref{sec:identitiesM}. They allow us to bring the expressions
for $c$ and $C_1/C_2$ to the following form:
\beq
c=-2g\theta+\frac{\det{\cal M}_{2L+1}^{(1,2)}}{\det{\cal M}_{2L+1}^{(1,1)}},\;\;
C_1/C_2=\frac{\det{\cal M}_{2L}^{(1,2)}}{\det{\cal M}_{2L}^{(1,1)}}.
\label{eq:Cnew}
\eeq
Finally we can plug \eqref{eq:Cnew} into \eqref{eq:BfromC} and write our main result for $\Gamma_L(g)$

\beq
\Gamma_L(g)=(\phi-\theta) g\left(r_{2L-1}-r_{2L}\right),\; r_N=\frac{\det {\cal M}^{(1,2)}_{N+1}}{\det {\cal M}_{N}}.
\label{eq:mainresult1}
\eeq
Using the identities given in Appendix \ref{sec:identitiesM}, we can represent it in a compact form. The final formula reads

\beq
\boxed{
\Gamma_L(g)=\frac{\phi-\theta}{4}\partial_\theta\log\frac{\det{\cal M}_{2L+1}}{\det{\cal M}_{2L-1}}
}\label{eq:mainresult}\ .
\eeq
This is our main result which was announced in the Introduction. As an example, for $L=1$ it reduces to
\beq
	\Gamma_1(g)=(\phi-\theta)g
	\ofrac{I_1^\theta}
	\frac{ \(I_2^\theta\)^3-2
   I_1^\theta I_2^\theta I_3^\theta
  +\(I_1^\theta\)^2 I_4^\theta}
  {
   \(I_1^\theta\)^2-I_1^\theta I_3^\theta
   +\(I_2^\theta\)^2},
\eeq
while for higher values of $L$ the expression becomes quite lengthy.

A form more suitable for some calculations is
\beq
\Gamma_L(g)=(-1)^{L+1}(\phi-\theta)g  \frac{\det{\cal M}_{2L+1}^{(1,2L+2)} }{\det{\cal M}_{2L}}.
\label{eq:mainresult2}
\eeq
Notice that here the matrix in the numerator is just ${\cal M}_{2L}$ with all indices of deformed Bessel functions $I_n^\theta$
increased by 1.

The explicit result for the energy \eq{eq:mainresult} concludes our analytical solution of the cusp TBA equations. In the next subsection we will describe several checks of the result.

\subsection{Weak and strong coupling limit}

While for $L=0$ our result matches fully the prediction from localization, at nonzero $L$ our result is new. Here we will show that it passes several nontrivial checks.

At strong coupling our computation should reproduce the energy of the corresponding classical string solution which was computed in \cite{Gromov:2012eu} (see also \cite{Drukker:2011za} for relevant calculations at strong and at weak coupling). To do this we first expanded the energy at large $g$ and fixed $L$ for several first values of $L$. The dependence on $L$ happened to be polynomial which allows us to easily extend the result to
an arbitrary $L$ (see \eqref{BpolL}):
\beqa
\nn
    \frac{\Gamma_L}{2(\phi-\theta)\theta}&=&
    -\frac{g}{\sqrt{\pi ^2-\theta ^2}}
    +\frac{6L+3}{8 \(\pi ^2- \theta ^2\)}
    -\frac{3\left( \(6 L^2+6 L+1\)\pi^2-{2 \theta ^2 }(L+1)L\right)}
   {128 g\pi^2\left(\pi ^2-\theta
   ^2\right)^{3/2}}+\dots \\
   \label{BpolL1}
\eeqa
To compare with the classical string energy we re-expanded this formula in the regime when $L$ and $g$ are both large, but  ${\cal L}=L/g$ is fixed. Then at leading order in $g$ we found (more details are given in Appendix \ref{sec:strong})
\beqa
\la{ELexmain}\frac{{\Gamma_L}}{2(\phi-\theta)\theta}\;&=&
\left(-\frac{g}{\pi }+\frac{3 L}{4 \pi ^2}-\frac{9 L^2}{64
   g\pi^3  }-\frac{5 L^3}{256  g^2 \pi^4  }+\frac{45 L^4}{16384 g^3 \pi
   ^5}\right)\\
\nn&+&\theta ^2
    \left(-\frac{g}{2 \pi ^3}+\frac{3 L}{4 \pi ^4}-\frac{21 L^2}{128 g  \pi^5  }-\frac{L^3}{16
    g^2 \pi^6  }-\frac{105 L^4}{32768  g^3 \pi^7
    }\right)\\
\nn&+&\theta ^4  \left(-\frac{3 g}{8 \pi
   ^5}+\frac{3 L}{4 \pi ^6}-\frac{99 L^2}{512   g\pi^7  }-\frac{3 L^3}{32  g^2 \pi^8
    }-\frac{2085 L^4}{131072  g^3 \pi^9  }\right)\\
\nn&+&\theta ^6  \left(-\frac{5 g}{16 \pi ^7}+\frac{3 L}{4 \pi
   ^8}-\frac{225 L^2}{1024  g \pi^9  }-\frac{L^3}{8  g^2 \pi^{10}  }-\frac{7905
   L^4}{262144  g^3 \pi^{11}  }\right)\\
\nn&+&\theta ^8  \left(-\frac{35 g}{128 \pi ^9}+\frac{3 L}{4 \pi ^{10}}-\frac{1995 L^2}{8192 g \pi^{11}}-\frac{5 L^3}{32  g^2 \pi^{12}  }-\frac{97425 L^4}{2097152  g^3 \pi^{13}
    }\right)\;,
\eeqa
which perfectly matches the expansion of the classical string energy from \cite{Gromov:2012eu}! Since the classical energy was derived without appealing to integrability, this matching is a direct test of our calculation for nonzero $L$.

At weak coupling we can compare our result to the leading Luscher correction to the energy.
 This correction was computed, as well as shown to follow from the TBA equations, in \cite{Correa:2012hh}, \cite{Drukker:2012de} for generic $\phi$ and $\theta$. It was also reproduced in \cite{Correa:2012nkNew}\footnote{except for the overall coefficient which was not fixed in \cite{Correa:2012nkNew}} by a direct perturbative calculation. When $\theta\sim\phi$ this Luscher correction reduces to
\beq
	\Gamma_L=(\phi-\theta)g^{2L+2}
	\frac{(-1)^L(4\pi)^{1+2L}}{(1+2L)!}B_{1+2L}\(\frac{\pi-\theta}{2\pi}\)+\cO(g^{2L+4})
\eeq
where $B_{1+2L}$ are the Bernoulli polynomials. For $L=0,1,2,3,4$ we have checked that this expression precisely coincides with the leading weak-coupling term of our result.



\section{Relation to the ${\bf P}\mu$-system}
\label{sec:Pmu}
In \cite{PmuPRL} a new efficient and compact formalism describing the spectrum of planar AdS/CFT
was proposed. It is based on the underlying (classical) integrability of the Hirota equation,
analytical properties of T-functions in some special gauge and an additional structure identified with
${\mathbb Z}_4$ symmetry observed in \cite{Gromov:2011cx}. In this section our goal is to identify the main objects
entering the ${\bf P}\mu$-system, i.e. a 4D vector function  ${\bf P}_a$ of the spectral parameter,
and an antisymmetric $4\times 4$ matrix $\mu_{ab}$. According to \cite{PmuPRL}
these objects have the following properties:
\begin{itemize}
\item ${\bf P}_a$ has a single cut $[-2g,2g]$ on the main ``physical sheet" and no other singularities. The analytical continuation under the cut denoted as $\tilde {\bf P}_a$ is given by
\beq\la{Pmu}
\tilde {\bf P}_a=-\mu_{ab}\chi^{bc}{\bf P}_c\;\;,\;\;\chi^{bc}=\(\begin{array}{cccc}0&0&0&-1\\0&0&1&0\\0&-1&0&0\\1&0&0&0\end{array}\),
\eeq
or, explicitly,

\begin{equation}
\begin{cases}
\tilde {\bf P}_1=-\mu_{1,2}{\bf P}_3+\mu_{1,3}{\bf P}_2-\mu_{1,4}{\bf P}_1,\\
\tilde {\bf P}_2=-\mu_{1,2}{\bf P}_4+\mu_{2,3}{\bf P}_2-\mu_{2,4}{\bf P}_1,\\
\tilde {\bf P}_3=-\mu_{1,3}{\bf P}_4+\mu_{2,3}{\bf P}_3-\mu_{3,4}{\bf P}_1,\\
\tilde {\bf P}_4=-\mu_{1,4}{\bf P}_4+\mu_{2,4}{\bf P}_3-\mu_{3,4}{\bf P}_2.\\
\end{cases}
\label{eq:pmusystem}
\end{equation}

\item As the branch points are of the square-root type, the analytical continuation of ${\tilde {\bf P}}_a$ under the cut should give ${\bf P}_a$
again which is the case if
\beq
\mu\chi\mu\chi={\bf 1}.
\eeq
\item $\mu_{ab}$ also has no poles, but infinitely many cuts $[-2g+i n,2g+in],\;n\in{\mathbb Z}$ which are all related to each other.
The discontinuity on the cut on the real axis (i.e. with $n=0$) is given by
\beq\la{muab}
\tilde\mu_{ab}-\mu_{ab}= {\bf P}_a \tilde{\bf P}_b- {\bf P}_b \tilde{\bf P}_a.
\eeq
\item $\mu$ is periodic as a function with ``long" cuts (i.e. connecting the branch points $\pm 2g$ with infinity, see \cite{Gromov:2011cx}), which means that
\beq\la{mush}
\mu_{ab}(u+i)=\tilde\mu_{ab}(u)\;.
\eeq
\item Any Y-function can be expressed solely in terms of ${\bf P}_{a}$ and $\mu_{ab}$. In particular
\beq\la{Y11Y22}
Y_{1,1}Y_{2,2}=1+\frac{{\bf P}_1\tilde{\bf P}_2-{\bf P}_2\tilde{\bf P}_1}{\mu_{12}}\;.
\eeq
\end{itemize}

Since we posses an exact solution of the Y-system we are in a rather unique situation when we can test the proposal thoroughly.
We will also check an important assumption made in the derivation above about the absence of cuts in $\eta^2$ and show that this is indeed a prediction of the ${\bf P}\mu$-system.

The way the construction goes is the following. We have to make a gauge transformation to a special ${\mathbb T}$-gauge
introduced in \cite{Gromov:2011cx} and then read ${\bf P}_1$ and ${\bf P}_2$ off ${\mathbb T}_{1,s}$:
\beq
{\mathbb T}_{1,s}={\bf P}_1^{[+s]}{\bf P}_2^{[-s]}-{\bf P}_2^{[+s]}{\bf P}_1^{[-s]}={\bf h}^{[+s]}{\bf h}^{[-s]}T_{1,s},
\label{eq:TP}
\eeq
where the function $\bf h$ defines the gauge transformation and should be found from the gauge-fixing condition \cite{Gromov:2011cx}\footnote{Namely, Eq. (5.36) in \cite{Gromov:2011cx}, where one should also use the identification
$\mu_{12}={\cal F}^+$}:
\beq
\frac{{\mathbb T}_{1,1}(u-\tfrac{i}{2}-i0){\mathbb T}_{1,1}(u+\tfrac{i}{2}+i0)-
{\mathbb T}_{1,1}(u-\tfrac{i}{2}+i0){\mathbb T}_{1,1}(u+\tfrac{i}{2}-i0)}{{\mathbb T}_{1,2}(u)}=
\mu_{12}\(Y_{1,1}Y_{2,2}-1\),
\eeq
which in our notation gives:
\beq
{\bf h}^{[+0]} {\bf h}^{[-0]}\frac{\sin\theta\rho}{\sin\theta}=\mu_{12}\(1-Y_{11}Y_{22}\)=	2\epsilon\(\Phi-\Psi\)\mu_{12}\;.
\eeq
We can already see an important feature of our limit which is that ${\bf h}\sim \epsilon^{1/2}\to 0$. This is the main reason for
the simplification in the construction as we shall see soon. Using \eqref{eq:FiNLIE1} we get
\beq
{\bf h}^{[+0]} {\bf h}^{[-0]}=-2\epsilon\eta\mu_{12}\;
\label{eq:hmu}.
\eeq
Comparing \eq{eq:TP} with our ansatz from section \ref{sec:FiNLIE} \footnote{notice that we are free to replace $Q$-functions by their linear combinations, since $T$ is given by a determinant}
\beq
{T}_{1,s}=\frac{Q_1^{[+s]}Q_2^{[-s]}-Q_1^{[-s]}Q_2^{[+s]}}{2i \sin\theta}\;\;,\;\;Q_1=e^{+\theta u-i \theta G}
\;\;,\;\;Q_2=e^{-\theta u+i \theta G}
\eeq
 we deduce

\beq\la{PhQ}
{\bf P}_1\propto{\bf h} Q_1\;\;,\;\;
{\bf P}_2\propto{\bf h} Q_2\;,
\eeq
which implies that ${\bf P}_1, {\bf P}_2$ are small in our limit.
Then due to \eq{muab} this means that the discontinuity of $\mu_{12}$
is small and thus to the leading order $\mu_{12}$ is a
meromorphic function without cuts. Then from \eq{Y11Y22} and $Y_{11}Y_{22}-1=2\epsilon\left( \Psi-\Phi\right)$ we see that $\mu_{12}$ must have zeros at
$u=i n/2$, because from the first Bremsstrahlung TBA equation, Eq. \eq{eq:TBAfinaleq1}, $\Psi-\Phi$ has poles at those points. Furthermore $\mu_{12}$ is a periodic function
due to \eq{mush}. We will see that we can identify $\mu_{12}=\sinh(2\pi u)$.
Indeed with this choice $\sinh(2\pi u)$ appearing in \eq{eq:hmu} through
\beq
\eta=\frac{\cos\theta}{\tilde C}\frac{2\pi u}{\sinh 2\pi u}\sqrt{\tilde {\bf Q}_-
\tilde {\bf Q}_+
 {\bf Q}_-
 {\bf Q}_+
 }
 \eeq
 cancels and we get
\beq\la{hheq}
{\bf h}^{[+0]}{\bf h}^{[-0]}=-\frac{4\pi u\epsilon \cos\theta }{\tilde C}
\sqrt{\tilde {\bf Q}_-
\tilde {\bf Q}_+
 {\bf Q}_-
 {\bf Q}_+
 }.
\eeq
Requiring ${\bf h}^2$ to be regular and to decay as $1/x^{2L+1}$ \cite{Gromov:2011cx} we get
from \eq{hheq}
\beq
{\bf h}^2=-\frac{4\pi\epsilon \cos\theta }{\tilde C}\frac{ u}{x^{2L+2}}\tilde {\bf Q}_-
\tilde {\bf Q}_+,
\eeq
which then implies for ${\bf P}_1$ and ${\bf P}_2$
\beq
{\bf P}_1= C\frac{\sqrt{u}e^{+\theta u}\tilde \bQ_-}{x^{L+1}},\;\;
{\bf P}_2=(-1)^{L+1} C\frac{\sqrt{u}e^{-\theta u}\tilde \bQ_+}{x^{L+1}},
\eeq
where $C=\sqrt{-2\pi i (\phi-\theta)/\tilde C}$.
Next we can see that upon identification $\mu_{14}=\mu_{23}=\mu_{34}=0$ and $\mu_{13}=-\mu_{24}=1$
the Baxter equation \eqref{eq:baxterconsider}, \eqref{eq:conjecture} becomes \eq{eq:pmusystem} if in addition we set
\beqa
{\bf P}_3&=&-C e^{g\theta (x-1/x)}\sqrt{u}P_L(+x),\\
{\bf P}_4&=&C e^{-g\theta (x-1/x)}\sqrt{u}P_L(-x)\;.
\eeqa
We can check now that all properties of the ${\bf P}\mu$ system are satisfied.

Now when we see that  ${\bf P}\mu$ works nicely in our case
we can revert the logic and start from ${\bf P}\mu$-system which
by itself can be derived starting from TBA equations in a rather nontrivial way
(see \cite{toappearPMu}). From \eq{hheq} and \eq{PhQ} and using that in our ansatz $Q_1Q_2={\rm const}$ we can
write $\eta$ up to a constant as
\beq\la{etamu}
\eta^2\propto\frac{{\bf P}_1{\bf P}_2 \tilde{\bf P}_1\tilde{\bf P}_2}{\mu^2_{12}}\;.
\eeq
From here we see that in general $\mu_{12}$ has infinitely many cuts and also $\tilde{\bf P}_a$
will have infinitely many cuts and thus most likely $\eta$ would in general have infinitely many cuts.
However, the particularity of our limit is that ${\bf P}_a\simeq \epsilon^{1/2}\to 0$
and thus the discontinuity of $\mu_{ab}$ is vanishing. As a result  $\tilde{\bf P}_a$ will have only
the cut on the real axis which obviously cancels in the combination \eq{etamu}.
Thus we conclude that $\eta^2$ indeed should have no cuts as was anticipated in the
previous sections. This completes our derivation.

\section{Conclusions}
\label{sec:Conclusions}
In this paper we have computed explicitly the generalized cusp anomalous dimension $\Gamma_L(g,\phi,\theta)$ in the near-BPS limit when $\phi\approx\theta$. We have thus extended the $\theta=0$ calculation of \cite{Gromov:2012eu} to the arbitrary $\theta$ case. Our result \eq{eq:mainresult} is fully non-perturbative and covers generic values for three ($g, \ L$ and $\theta$) out of four parameters in the cusp anomalous dimension.

At $L=0$ our result matches an earlier localization calculation. For nonzero $L$ it serves as a new integrability-based prediction for localization techniques, and is fully confirmed by nontrivial checks both at strong and at weak coupling.

To compute the cusp anomalous dimension we solved exactly the system of near-BPS cusp TBA equations. The angle $\theta$ enters these equations as a twist parameter similar to those arising for integrable $\gamma$ and $q-$deformations of $\cN=4$ SYM.\footnote{See e.g. \cite{deformed} for a discussion of the spectral TBA and Y-system in the deformed theory.} We have generalized the powerful FiNLIE approach to the twisted case and thus reduced the infinite TBA system to a finite set of equations. We then solved these equations analytically, encountering deformed versions of several remarkable structures (effective Bethe equations/Baxter equations) discovered at $\theta=0$ in \cite{Gromov:2012eu}. The very rare situation of having an exact solution of the TBA also allowed us to test a novel ${\bf P}\mu$-system reformulation of the spectral problem.

\bigskip

Our result for $\Gamma_L$ has a form of a logarithmic derivative of a ratio of determinants, which hints that it could be obtained as an expectation value of some quantity in a matrix model. As in the $\theta=0$ case \cite{Gromov:2012eu} we expect that matrix model techniques should be very useful to analyze the semiclassical expansion of our predictions at large $L$.

It would be interesting to explore further the role of the effective Bethe equations that arose in our construction, and to study their interplay with the Bethe ansatz for excitations on top of the $Z^L$ insertion in the Wilson loop.

While in this paper we have solved the generic near-BPS case, other regimes may be found where the cusp TBA would also admit an explicit solution -- for example, the ``ladders'' limit when $i\theta\to\infty$ with fixed $ge^{i\theta}$. On the whole, an impressive amount of perturbative data for the cusp anomalous dimension is available \cite{Correa:2012nkNew, BykZar, Henn:2013wfa}, which should be reproduced from the TBA.

In general, many of the open questions discussed in \cite{Gromov:2012eu} should be easier to attack now that we have an extra parameter $\theta$ in the exact solution. We hope to return to some of these points in future work.

\section*{Acknowledgements}
We are grateful to J. Henn, D. Volin, A. Sever and S. Valatka for helpful discussions and comments.
The work of F.L.-M. was supported in part by grants RFBR-12-02-00351-a, PICS-12-02-91052, and by Russian Ministry of Science and Education under the grant 2012-1.1-12-000-1011-016 (contract number 8410).
The authors acknowledge the support under the GATIS network; the research leading to these results has received funding from the People Programme (Marie Curie Actions) of the European Union's Seventh Framework Programme FP7/2007-2013/ under REA Grant Agreement No 317089.

\appendix

\section{Notation and conventions}
\label{sec:notationandconventions}
In this appendix we summarized the notation which is used throughout the paper. The basic definitions are in the first subsection, and the second one contains a glossary of integration kernels.
\subsection*{Basic notation}
\label{sec:notation}

\beq
 f^{[\pm a]}\equiv f(u\pm ia),\ \ f^{\pm} \equiv f(u\pm i/2),
\eeq
\beq
f^{[\pm0]}=f(u\pm i0),\ \ f^{\pm_\pm}=f(u\pm i/2\pm i0).
\eeq
\beq
I_{m,n}\equiv\delta_{m+1,n}+\delta_{m-1,n}.
\eeq
The Zhukovsky transformation $x(u)$ is defined by
\beq
\frac{u}{g}=x(u)+\ofrac{x(u)}.
\eeq
with $g=\frac{\sqrt{\lambda}}{4\pi}$.
We also found it convenient to denote
\begin{align}
T=e^{i\theta}, \  c_a=e^{2 i G(ia/2)}, \ y_a=x(ia/2),
\end{align}
where $G$ is the resolvent from \eq{eq:Gdefinition}.

\subsection*{Kernels in the TBA}
\label{sec:Kernels}
We denote by $*$ the convolution over the full real axis from $-\infty$ to $\infty$, and by $\hat *$ the convolution over the range $-2g<u<2g$.

Our definitions of the kernels coincide with the ones used in \cite{Correa:2012hh} and \cite{Gromov:2012eu}, and we summarize them below:
\begin{align}
{\mathbf s}(u,v)=\frac{1}{2 \cosh(\pi (u-v))},
\end{align}

\beq
	K_a(u,v)=\frac{2a}{\pi(a^2+4(u-v)^2)},
	\label{eq:kernelsK}
\eeq

\beq
	\ \hat K_a(u)=\hat K_{y,a}(u,0)=\sqrt{\frac{4g^2-u^2}{4g^2+a^2/4}}K_a(u),
	\ \tilde 	K_a(u)=\sqrt{\frac{4g^2+a^2/4}{4g^2-u^2}}K_a(u),
	\label{eq:kernelsKhattilde}
\eeq

\begin{align}
& K_{n,m}(u,v)=\sum^{\frac{n-1}{2}}_{j=-\frac{n-1}{2}}\sum^{\frac{m-1}{2}}_{k=-\frac{m-1}{2}}K_{2j+2k+2}(u,v),\\
& K(u,v)=\frac{1}{2\pi i}\sqrt{\frac{4g^2-u^2}{4g^2-v^2}}\frac{1}{v-u},
\end{align}

\beq
\log F_a(a,g)=\left.\tilde K_a\hat *\log\frac{\sinh(2\pi u)}{2\pi u}\right|_{u=0}.
\eeq

\begin{align}
& r(u,v)=\frac{x(u)-x(v)}{\sqrt{x(v)}},\  b(u,v)=\frac{1/x(u)-x(v)}{\sqrt{x(v)}},\\
&{\cal R}^{(ab)}_{nm}=\sum^{\frac{n-1}{2}}_{j=-\frac{n-1}{2}}\sum^{\frac{m-1}{2}}_{k=-\frac{m-1}{2}}\frac{1}{2\pi i}\frac{d}{dv}\log\frac{r(u+ia/2+ij,v-ib/2+ik)}{r(u-ia/2+ij,v+ib/2+ik)},\\
&{\cal B}^{(ab)}_{nm}=\sum^{\frac{n-1}{2}}_{j=-\frac{n-1}{2}}\sum^{\frac{m-1}{2}}_{k=-\frac{m-1}{2}}\frac{1}{2\pi i}\frac{d}{dv}\log\frac{b(u+ia/2+ij,v-ib/2+ik)}{b(u-ia/2+ij,v+ib/2+ik)},
\end{align}

Given the definitions above one can prove the following identities (see \cite{Correa:2012hh}):

\begin{align}
&{\cal R}^{(10)}_{a1}(u,v)+{\cal B}_{a1}^{(10)}(u,v)=K_a(u,v),\\
&{\cal R}^{(10)}_{a1}(u,v)-{\cal B}_{a1}^{(10)}(u,v)=K(u+ia/2,v)-K(u-ia/2,v),\\
&{\cal R}^{(01)}_{1a}(u,v)+{\cal B}^{(01)}_{1a}(u,v)=K_a(u,v),\\
&{\cal R}^{(01)}_{1a}(u,v)-{\cal B}^{(01)}_{1a}(u,v)=\hat K_{y,a}(u,v)=K(u,v-ia/2)-K(u,v+ia/2),\\
&{\cal R}_{2n}^{(01)}=\frac{1}{2}\left(\hat K_n^+-\hat K_n^-+K_n^++K_n^-\right)
\end{align}
\beqa
\tilde K_{ab}&=&{\cal R}_{ab}^{(10)}+{\cal B}_{ab-2}^{(10)}=\\ \nn
&=&\frac{1}{2}\left(\tilde K_a^{[b-1]}-\tilde K_a^{[-b+1]}+K_a^{[b-1]}+K_a^{[-b+1]}\right)+\sum\limits_{r=1}^{a}K_{b-a-3+2r}\\
\hat K_{ba}&=&{\cal R}_{ba}^{(01)}+{\cal B}_{b-2,a}^{(01)}=\\ \nn
&=&\frac{1}{2}\left(\hat K_a^{[b-1]}-\hat K_a^{[-b+1]}+K_a^{[b-1]}+K_a^{[-b+1]}\right)+\sum\limits_{r=1}^{a}K_{b-a-3+2r}
\eeqa

\section{Bremsstrahlung TBA in the near-BPS limit}
\label{sec:BremsstrahlungApp}
Here we give more details concerning the derivation of the simplified Bremsstrahlung TBA system from the cusp TBA equations of \cite{Correa:2012hh,Drukker:2012de}.

\subsection*{Asymptotic solutions}
\label{sec:Asymptotics}
As it was mentioned in section \ref{sec:TBA}, the main difference between the spectral TBA and the cusp TBA is the asymptotic large $L$ solution. In order to obtain the asymptotic solution in our limit we expand in small $\epsilon$ the asymptotic solution given in \cite{Correa:2012hh} (denoting the asymptotic Y-functions by bold font as in \cite{Correa:2012hh})
\begin{align}
	&\textbf{Y}_{1,1}=1/{\textbf{Y}}_{2,2}=-\frac{\cos\theta}{\cos\phi}\approx-1-\epsilon,\\
	&\textbf{Y}_{a,1}=\frac{\sin^2\phi}{\sin(a+1)\phi\sin(a-1)\phi}\approx A_a-\frac{\epsilon}{\tan\phi_0} B_a,
	\label{eq:Ya1asymptot}
 \\
	&1/{\textbf{Y}}_{1,s}=\frac{\sin^2\theta}{\sin(s+1)\theta\sin(s-1)\theta}\approx A_s+\frac{\epsilon}{\tan\phi_0} B_s,
\label{eq:Ybolda0}
\end{align}
with $A_a$ and $B_a$ given by
\beq
	A_a=\frac{\sin^2\phi_0}{\sin(1+a)\phi_0\sin(a-1)\phi_0},\
	B_a=\frac{4\sin\phi_0\sin a\phi_0(a\cos a\phi_0\sin\phi_0-\cos\phi_0\sin{a\phi_0})}
	{(\cos 2\phi_0-\cos 2a\phi_0)^2}.
\eeq
Thus for the asymptotic solution the leading orders of the Y-functions (as defined in \eq{yexp1}) read
\begin{align}
&\mathbf{\Phi}=\mathbf{\Psi}=\frac{1}{2},\\
&\mathbf{{\cal Y}_a}=A_a, \\
&\mathbf{{\cal X}_a}=\frac{B_a\cot\phi_0}{A_a},
\end{align}
and we also have
\begin{align}
\textbf{Y}_{a,0}\approx\frac{((\phi-\theta) \sin{a\phi_0)}^2}{u^2}a^2\left(F(a,g)\frac{z_0^{[-a]}}{z_0^{[+a]}}\right)^2,\  u\rightarrow 0.
\end{align}

\subsection*{Derivation of Bremsstrahlung TBA}

First, plugging into the cusp TBA equations of \cite{Correa:2012hh} our expansion \eqref{yexp1} of the Y-functions and using the asymptotic solutions derived in the previous section we obtain
\begin{align}
\Psi-\frac{1}{2}&=K_{m-1}*\left({\cal X}_m\frac{{\cal Y}_m}{1+{\cal Y}_m}-\frac{B_m\cot\phi_0}{1+A_m}\right)-\pi \mathbb{C}_a{\cal R}_{1a}^{(01)}(u,0),\\
\Phi-\frac{1}{2}&=K_{m-1}*\left({\cal X}_m\frac{{\cal Y}_m}{1+{\cal Y}_m}-\frac{B_m\cot\phi_0}{1+A_m}\right)-\pi \mathbb{C}_a{\cal B}_{1a}^{(01)}(u,0),\\
\log\frac{{\cal Y}_m}{A_m}&=-K_{m-1,n-1}*\log\frac{1+{\cal Y}_n}{1+A_n}-K_{m-1}\hat{*}\log\frac{\Psi}{\Phi},\\
{\cal X}_m-\frac{B_m\cot\phi_0}{A_m}&=-K_{m-1,n-1}*\left(\frac{{\cal Y}_n }{1+{\cal Y}_n}{\cal X}_n-\frac{ B_n\cot\phi_0}{1+A_n}\right)
+\pi {\mathbb C}_n\left[{\cal R}^{(01)}_{m n}+{\cal B}^{(01)}_{m-2, n}\right](u,0),\\
\Delta_a&=[{\cal R}_{ab}^{10}+{\cal B}_{a,b-2}^{10}]\hat *\log\frac{1+{\cal Y}_b}{1+A_b}+{\cal R}_{a1}^{10}\hat *\log\left(\frac{\Psi}{1/2}\right)-
{\cal B}_{a1}^{10}\hat *\log\left(\frac{\Phi}{1/2}\right),\\
\mathbb{C}_a&=(-1)^{a+1} a \frac{\sin a\phi_0}{\tan\phi_0}\left(\sqrt{1+\frac{a^2}{16g^2}}-\frac{a}{4g}\right)^{2+2L}  F(a,g)e^{\Delta_a}.
\end{align}
Using the strategy described for the small angles case in Appendix F of \cite{Correa:2012hh} we can simplify these equations and get
\begin{align}
&\Psi-\Phi=\pi \mathbb{C}_a \left[{\cal B}^{(01)}_{1 a}-{\cal R}^{(01)}_{1 a}\right](u,0),\\
&\Psi+\Phi=\left(\frac{B_2\cot\phi_0}{A_2(1+A_2)}+1\right)-2{\mathbf s}*\frac{{\cal X}_2}{1+{\cal Y}_2}+2\pi \mathbb{C}_n \textbf{s}*{\cal R}_{2n}^{(01)}-\pi \mathbb{C}_a\left[{\cal R}^{(01)}_{1 a}+{\cal B}^{(01)}_{1 a}\right](u,0),\\
&\log \frac{{\cal Y}_m}{A_m}={\mathbf s}*I_{m,n}\log\left(\frac{{\cal Y}_n }{1+{\cal Y}_n}\frac{1+A_n}{A_n}\right)+\delta_{m,2}\textbf{s}\hat{*}\log\frac{\Phi}{\Psi},\\
&{\cal X}_m=\frac{B_m\cot\phi_0}{A_m}+{\mathbf s}* I_{m,n}\left(\frac{{\cal X}_n }{1+{\cal Y}_n}-\frac{B_n\cot\phi_0}{(1+A_n)A_n}\right)+\pi \mathbb{C}_m\mathbf{s}+\delta_{m,2}\mathbf{s}\hat{*}(\Phi-\Psi).
\end{align}
Finally, substituting the explicit form of $A_n,B_n$ we can simplify the equations even further. Using that
\begin{align}
\frac{B_2\cot\phi_0}{A_2(1+A_2)}=-1,
\end{align}
\begin{align}
\frac{B_m\cot\phi_0}{A_m}-\frac{1}{2}I_{m,n}\frac{B_n/A_n\cot\phi_0}{1+A_n}=0,
\end{align}
and
\begin{align}
\frac{1}{2}\log\frac{1+A_{m+1}}{A_{m+1}}\frac{1+A_{m-1}}{A_{m-1}}+\log A_m=0.
\end{align}
we obtain the final form of the equations as written in section \ref{sec:TBA}.

\section{Derivation of FiNLIE -- details}
\label{sec:DerivationofFiNLIE}
In this appendix we extend section \ref{sec:FiNLIE} by explaining in more detail reduction of the Bremsstrahlung TBA equation \eqref{eq:TBAfinaleq1}-\eqref{eq:TBAfinaleq6} to a set of three equations \eqref{eq:FiNLIE1}-\eqref{eq:FiNLIE3} for the quantities $\eta(u),\rho(u),\mathbb{C}_a$ called FiNLIE.

\subsection*{$\Psi\pm\Phi$ equations}

The left hand sides of the first two Bremsstrahlung TBA equations are $\Psi\pm\Phi$. Let us express this combination in terms of $\rho$ and $\eta$.
Using the definition \eqref{eq:etadef} of $\eta$ and the explicit form \eqref{eq:Texplicit0} of ${\cal T}_m$ we get
\begin{align}
\Psi-\Phi=\frac{{\cal T}_1^{-_+}{\cal T}_1^{+_-}-{\cal T}_1^{-_-}{\cal T}_1^{+_+}}{{\cal T}_2}\eta=\frac{\sin{\rho\theta}}{\sin\theta}\eta
\label{eq:PsiminusPhi}
\end{align}
and
\begin{align}
\Psi+\Phi=\frac{{\cal T}^{+_-}_1{\cal T}^{-_+}_1+{\cal T}^{-_-}_1{\cal T}^{+_+}_1}{{\cal T}_2 }\eta=
\frac{\cos{\rho\theta}\cos{(2-G^++G^-)\theta}-\cos{(2\slashed{G}-G^+-G^-)\theta}}{\sin{\theta}\sin{(2-G^++G^-)\theta}}\eta.
\label{eq:PsiplusPhi}
\end{align}
Comparing this with the first two equations of Bremsstrahlung TBA (Eq. \eq{eq:TBAfinaleq1} and Eq. \eq{eq:TBAfinaleq2}) gives the first two FiNLIE equations \eq{eq:FiNLIE1} and \eq{eq:FiNLIE2}.

\subsection*{Equation for $\Delta_a$}
Here we discuss the reduction of the equation \eqref{eq:TBAfinaleq5} for $\Delta_a$ to the third FiNLIE equation \eqref{eq:FiNLIE3}. Using identities for kernels \eqref{eq:TBAfinaleq5} can be written as
\begin{align}
\Delta_a=\frac{1}{2} K_a\hat *\log{\frac{\Psi}{\Phi}}+\frac{1}{2}\tilde K_a\hat *\log\left(\frac{\Psi\Phi}{1/4}\right)+\sum\limits_{b=2}^{\infty} \tilde K_{ab}*\log{\left(\frac{1+{\cal Y}_a}{1+A_a}\right)}.
\label{eq:Deltaa}
\end{align}
Let us introduce a notation for the asymptotic large $u$ values of ${\cal T}_a$:
\beq
\psi_a=\frac{\sin a\phi_0}{\sin\phi_0}.
\eeq
We can transform equation \eqref{eq:Deltaa} performing the same manipulations as in section 3.3 of \cite{Gromov:2012eu}.
The only difference is that we are using ${\cal T}_a$ divided by their asymptotic values $\psi_a$ in order to ensure validity of manipulations with infinite sums below. We express Y-functions through T-functions and split the infinite sums as follows:
\begin{align}
\sum\limits^\infty_{b=2}\tilde K_{ab}*\log\left(\frac{1+{\cal Y}_b}{1+A_b}\right)=\sum\limits_{b=2}^\infty \tilde K_{ab}*\left[\log\frac{{\cal T}^+_b}{\psi_b^+}+\log\frac{{\cal T}^-_b}{\psi_b^-}-\log\frac{{\cal T}_{b+1}}{\psi_{b+1}}-\log\frac{{\cal T}_{b-1}}{\psi_{b-1}}\right]=\\
=\sum\limits_{b=2}^\infty\left[\tilde K^+_{ab}+\tilde K^-_{ab}-\tilde K_{a,b-1}- \tilde K_{a,b+1}\right]*\log\frac{{\cal T}_b}{\psi_b}+\tilde K_{a,1}*\log\frac{{\cal T}_2}{\psi_2}-\tilde K_{a,2}*\log\frac{{\cal T}_1}{\psi_1}.
\label{eq:sumtransform}
\end{align}
We have checked numerically that this last shifting of indices in the infinite sums is valid, i.e.
\begin{align}
\lim\limits_{B\rightarrow \infty} \tilde K_{a B+1}*\log \frac{{\cal T}_B}{\psi_B}-\tilde K_{a B}*\log \frac{{\cal T}_{B+1}}{\psi_{B+1}}=0.
\end{align}
The expression in square brackets in \eqref{eq:sumtransform} is zero almost everywhere, as one can see from the equation
(46) of \cite{Gromov:2012eu}. Taking into account that $\tilde K_{a,1}=0$ one gets
\begin{align}
\Delta_a=\frac{1}{2}\tilde K_a\hat *\log\frac{\Psi\Phi{\cal T}_2^2}{{\cal T}^{+_-}_1{\cal T}^{+_+}_1{\cal T}^{-_-}_1{\cal T}^{-_+}_1}-\tilde K_a\hat *\log\frac{\psi_2}{\psi_1^+\psi_1^-}+\log\frac{2{\cal T}_a}{\psi_a}.
\end{align}
Recalling the definition of $\eta$ (the second equality is due to \eqref{eq:PhioverPsi1})
\begin{align}
\eta\equiv \frac{\Psi {\cal T}_2}{{\cal T}_1^{-_+}{\cal T}_1^{+_-}}=\frac{\Phi {\cal T}_2}{{\cal T}_1^{-_-}{\cal T}_1^{+_+}}
\label{eq:etadefApp}
\end{align}
we find
\begin{align}
\Delta_a=\tilde K_a \hat *\log\eta-\tilde K_a\hat *\log\frac{\psi_2}{\psi_1^+\psi_1^-}+\log\frac{2{\cal T}_a}{\psi_a}
\end{align}
Substituting the explicit form of $\psi_a$ and taking into account that $\tilde K_a\hat *1=1$ we finally obtain
\begin{align}
\Delta_a=\tilde K_a \hat *\log\eta+\left.\log\frac{{\cal T}_a}{\sin a\phi_0\cot\phi_0}\right|_{u=0}
\label{eq:etaeqApp}
\end{align}
thus giving Eq. \eq{eq:etaeq} presented in the main text.

\subsection*{Fixing the residues $b_a$}
\label{sec:fixba}

The residues $b_a$ of $\cG(u)$ at $ia/2$ satisfy a recursion relation which we derive in this section by comparing poles on both sides of \eqref{eq:TBAfinaleq3}. This recursion relation is used to find the residues of $\eta$ and obtain a relation between $G(ia/a)$ and $\rho(ia/2)$ which is described in section \ref{sec:relation} and in more details in appendix \ref{sec:FixingApp}.

By construction, the only poles in $1/Y_{1,m}(u)$ can originate from the poles of the resolvent at $u=ia/2,\  a\in \mathbb{Z}$. Consistently with this the equation \eqref{eq:TBAfinaleq3} tells the residue at $u=0$ should cancel and the residue at $u=i/2$ should obey

\beq
\log Y_{1,m}\approx-\epsilon\frac{\mathbb{C}_m}{2i(u-i/2)}.
\eeq
Expressing $Y_{1,m}$ through $T$-functions which are written in terms of the resolvent (see \eq{YthroughTfull}, \eq{eq:Texplicit}), and expanding at $u=i/2$, we obtain the
recursion relation for $b_a$ which was given in \eq{eq:recrelation}:
\beq
q_a b_{a-2}-(q_a+p_a)b_a+p_a b_{a+2}= \mathbb{C}_a,
\label{eq:recrelationApp}
\eeq
where explicitly
\begin{align}
&q_a=i\frac{4T^2c_{a-2}\log{T}(T^{2a}-c_a^2) }{(T^2c_{a-2}-c_a)(T^{2a}-T^2c_{a-2}c_a)},\\
&p_a=i\frac{4T^2c_{a+2}\log{T}(T^{2a}-c_a^2) }{(T^2c_{a}-c_{a+2})(T^{2a+2}-T^2c_{a}c_{a+2})},\\
&T=e^{i\theta},\\
&c_a=e^{2i G(ia/2)}.
\end{align}

\section{Fixing the residues of $\eta$ -- details}
\label{sec:FixingApp}
In this appendix we find the residues of $\eta$ at $ia/2$ from the second FiNLIE equation \eq{eq:FiNLIE2}. These residues are then used in the section \ref{sec:relation} to derive the relation \eqref{eq:densityequation}.

 To use \eq{eq:FiNLIE2} let us first of all get rid of the convolution in the r.h.s. by using
 the following property of  $\mathbf{s}$: for any function $f$ analytical in the strip $|\text{Im}\ u|<1/2$ but having poles at $u=\pm i/2$ with residues $\mp i \mathbb{C}/2$
\beq
f=\mathbf{s}*g+\pi \mathbb{C}\mathbf{s}(u)\ \Leftrightarrow\  f^+(u-i0)+ f^-(u+i0)= g(u).
\label{eq:invtranslationdeformed}
\eeq
Thus Eq. \eq{eq:FiNLIE2} takes the form
\beqa
\left(\frac{\cos{\rho\theta}\cos{(2-G^++G^-)\theta}-\cos{(2\slashed{G}-G^+-G^-)\theta}}{\sin{\theta}\sin{(2-G^++G^-)\theta}}\eta\right)^{+_-}+c.c=\nn&&\\
=-2\frac{ {\cal X}_2 }{1+{\cal Y}_2 }+\pi(\hat K_a^+-\hat K_a^-)\mathbb{C}_a. &&
\label{eq:bigeq}
\eeqa
Consider the residue at $ia/2$ of both sides of \eqref{eq:bigeq}. The terms that appear
after expanding the right hand side are proportional either to $\eta\sin\theta\rho$, or to $\eta\cos\theta\rho$. We know the residues of the first type of terms from the first FiNLIE equation \eq{eq:FiNLIE1}:
\beq
\underset{u=i a/2}{\text{Res }} \left(\eta\sin\theta\rho\right)=-\frac{\pi \mathbb{C}_a}{2\pi i}\sin\theta.
\eeq
To deal with the terms proportional to $\eta\cos\theta\rho$ let us introduce the notation
\begin{align}
\underset{u=i a/2}{\text{Res }} \left(\eta\cos\theta\rho\right)=\frac{e_a}{2\pi i}.
\end{align}
In the right hand side of \eqref{eq:bigeq} residues of ${\cal X}_a$ are expressed in terms of coefficients $b_a$ whose explicit value we do not know. However, a nice cancellation helps us to proceed. The residue at $ia/2$ of the ${\cal X}_m/(1+{\cal Y}_m)$ term has the form $k_1 b_{a-3}+k_2 b_{a-1}+k_3 b_{a+1}+k_4 b_{a+3}$, where $k_i$ are some clumsy coefficients. Nevertheless when we use the recursion relation
\eqref{eq:recrelation} to exclude $b_{a-3}$ and $b_{a+3}$ we see that $b_{a-1}$ and $b_{a+1}$ also cancel out! Thus we get an equation completely in terms of $e_a$ and $\mathbb{C}_a$:

\begin{multline}
-\frac{ T(T^2 c_{a-1}-c_{a+1})}{\pi (T^2-1)}\left(\frac{e_{a-1}(T^2 c_{a-3}-c_{a-1})}{c_{a-1}(T^4c_{a-3}-c_{a+1})}+
\frac{e_{a+1} (T^2c_{a+1}-c_{a+3})}{c_{a+1}(T^4 c_{a-1}-c_{a+3})}\right)+\\
+\frac{\mathbb{C}_{a+1}(c_{a+1}-T^2c_{a-1})(c_{a+1}^2+T^{2a+2})(T^2c_{a+1}-c_{a+3}) }{2c_{a+1}(T^{2a+2}-c_{a+1}^2)(T^4 c_{a-1}-c_{a+3})}+\\
+\frac{\mathbb{C}_{a-1}(c_{a-1}-T^2c_{a-3})(c_{a-1}^2+T^{2a-2})(T^2c_{a-1}-c_{a-3}) }{2c_{a-1}(T^{2a-2}-c_{a-1}^2)(T^4 c_{a-3}-c_{a+1})}=0.
\label{eq:hugeeq}
\end{multline}
One can see that this equation is solved by
\beq
e_a=\frac{\pi (T^2-1)}{2T}\mathbb{C}_a\frac{c_a^2+T^{2a}}{c_a^2-T^{2a}}.
\label{eq:solutioneaApp}
\eeq
By the same argument as in \cite{Gromov:2012eu} the initial conditions will help us to exclude other solutions. From \eqref{eq:FiNLIE1} one can see that $e_0=0$ (as $\rho(0)\ne 0$ in general), and $\mathbb{C}_0$ can be set to zero because the sum starts at $a=1$. Thus from \eqref{eq:hugeeq} it follows that \eqref{eq:solutioneaApp} holds for all even $a$. In order to fix $e_a$ at odd $a$ we look at the residue of the second FiNLIE equation \eqref{eq:FiNLIE2} at $u=i/2$. The only source of singularities in the right hand side are the terms with delta-function. Hence

\beq
e_1=\frac{\pi (T^2-1)}{2T}\mathbb{C}_a\frac{c_1^2+T^{2}}{c_1^2-T^{2}}.
\eeq
This agrees with \eqref{eq:solutioneaApp}, so our solution holds for all $a$.

\section{Strong coupling expansion }
\label{sec:strong}
Here we discuss the strong coupling expansion of the cusp anomalous dimension. In order to recover for $\theta=0$ the expansion given in \cite{Gromov:2012eu} (Appendix F) it is convenient to introduce the ``Bremsstrahlung function'' $B_L(g)$ related to the cusp anomalous dimension as
\beq
	\Gamma_L(g)=-2(\phi-\theta)\tan\theta B_L(g),
\eeq
It is straightforward to expand our result at strong coupling for fixed values of $L$, and we find, e.g.,
\beqa
	\nn
	\ofrac{\theta\cot\theta}B_{L=0}&=&\frac{g}{\sqrt{\pi ^2-\theta
   ^2}}-\frac{3}{8 \left(\pi ^2-\theta
   ^2\right)}+\frac{3}{128 g \left(\pi
   ^2-\theta ^2\right)^{3/2}}+\frac{3}{512
   g^2 \left(\pi ^2-\theta
   ^2\right)^2}
   \\
   &+&\frac{63}{32768 g^3
   \left(\pi ^2-\theta
   ^2\right)^{5/2}}
   +\mathcal{O}\(1/g^4\)\;,
   \\ \nn
   \ofrac{\theta\cot\theta}B_{L=1}&=&
   \frac{g}{\sqrt{\pi ^2-\theta
   ^2}}-\frac{9}{8 \left(\pi ^2-\theta
   ^2\right)}+\frac{3 \left(13 \pi ^2-4
   \theta ^2\right)}{128 \pi ^2 g
   \left(\pi ^2-\theta
   ^2\right)^{3/2}}\\ \nn
   &+&\frac{3 \left(-6
   \theta ^4+12 \pi ^2 \theta ^2+13 \pi
   ^4\right)}{512 \pi ^4 g^2 \left(\pi
   ^2-\theta ^2\right)^2}\\
   &+&\frac{9
   \left(-48 \theta ^6+64 \pi ^2 \theta
   ^4+136 \pi ^4 \theta ^2+31 \pi
   ^6\right)}{32768 \pi ^6 g^3 \left(\pi
   ^2-\theta
   ^2\right)^{5/2}}
   +\mathcal{O}\(1/g^4\)
\eeqa
We have computed such expansions for $L=0,1,\dots,4$ and when $\theta=0$ they reproduce the results in (195) of \cite{Gromov:2012eu}. As in \cite{Gromov:2012eu} we observed that the coefficients are polynomial in $L$, so we can now extrapolate to arbitrary $L$ which gives
\beqa
\label{BpolL}
	\ofrac{\theta\cot\theta}{B_L}&=&
	\frac{g}{\sqrt{\pi ^2-\theta ^2}}
	-\frac{6L+3}{8 \(\pi ^2- \theta ^2\)}
	+\frac{3\left( \(6 L^2+6 L+1\)\pi^2-{2 \theta ^2 }(L+1)L\right)}
   {128 g\pi^2\left(\pi ^2-\theta
   ^2\right)^{3/2}}
   \\ \nn
   &+&\frac{f_1}{512 g^2 \pi ^4
   \left(\pi ^2-\theta
   ^2\right)^2}
   -\frac{
   f_2}
   {32768\pi ^6 g^3 \left(\pi ^2-\theta
   ^2\right)^{5/2}}
   +\mathcal{O}\(1/g^4\)
\eeqa
where
\beqa
	f_1&=&-3 \theta ^4 L
   \left(2 L^2+3 L+1\right)+6 \pi ^2
   \theta ^2 L \left(2 L^2+3
   L+1\right)
   \\ \nn &&+\pi ^4 \left(10 L^3+15
   L^2+11 L+3\right)\;,\\
	f_2&=&18 \theta ^6 L \left(5 L^3+10 L^2+7
   L+2\right)-18 \pi ^2 \theta ^4 L \left(5
   L^3+10 L^2+11 L+6\right)\\ \nn
   &&-6 \pi ^4
   \theta ^2 L \left(55 L^3+110 L^2+47
   L-8\right)+9 \pi ^6 \left(10 L^4+20
   L^3-22 L^2-32 L-7\right)
\eeqa
Notice that for $\theta=0$ our expansion \eq{BpolL} reduces to that in Eq. (196) of \cite{Gromov:2012eu}.

We can now make a comparison with the classical string energy. Expanding \eq{BpolL} at fixed $\mathfrak{L}=L/g$ and large $g$, we get an expansion of the form
\beq
	\Gamma_L(g)=gE^{cl}\(\mathfrak{L}\)+E^{1-loop}\(\mathfrak{L}\)+\ofrac{g}E^{2-loop}\(\mathfrak{L}\)+\dots
	\, .
\eeq
The first term, $gE^{cl}$, is proportional to $\sqrt{\lambda}$ and is expected to reproduce the energy of the classical string configuration. Indeed, we found
\beqa
\la{ELex}\frac{g}{2(\phi-\theta)\theta}{E^{cl}}&=&
\left(-\frac{g}{\pi }+\frac{3 L}{4 \pi ^2}-\frac{9 L^2}{64
   g\pi^3  }-\frac{5 L^3}{256  g^2 \pi^4  }+\frac{45 L^4}{16384 g^3 \pi
   ^5}\right)\\
\nn&+&\theta ^2
    \left(-\frac{g}{2 \pi ^3}+\frac{3 L}{4 \pi ^4}-\frac{21 L^2}{128 g  \pi^5  }-\frac{L^3}{16
    g^2 \pi^6  }-\frac{105 L^4}{32768  g^3 \pi^7
    }\right)\\
\nn&+&\theta ^4  \left(-\frac{3 g}{8 \pi
   ^5}+\frac{3 L}{4 \pi ^6}-\frac{99 L^2}{512   g\pi^7  }-\frac{3 L^3}{32  g^2 \pi^8
    }-\frac{2085 L^4}{131072  g^3 \pi^9  }\right)\\
\nn&+&\theta ^6  \left(-\frac{5 g}{16 \pi ^7}+\frac{3 L}{4 \pi
   ^8}-\frac{225 L^2}{1024  g \pi^9  }-\frac{L^3}{8  g^2 \pi^{10}  }-\frac{7905
   L^4}{262144  g^3 \pi^{11}  }\right)\\
\nn&+&\theta ^8  \left(-\frac{35 g}{128 \pi ^9}+\frac{3 L}{4 \pi ^{10}}-\frac{1995 L^2}{8192 g \pi^{11}}-\frac{5 L^3}{32  g^2 \pi^{12}  }-\frac{97425 L^4}{2097152  g^3 \pi^{13}
    }\right)\;.
\eeqa
and all coefficients here match perfectly the  expansion of the classical string energy from Eq. (193) of \cite{Gromov:2012eu} ! This is a deep test of our computation at $L\neq 0$ against a result which does not rely on integrability.

\section{Identities for ${\cal M}_N$}
\label{sec:identitiesM}
In this appendix we describe the determinant identities which, in particular, allow us to switch between different representations \eqref{eq:mainresult1}, \eqref{eq:mainresult}, \eqref{eq:mainresult2} of the final result. Though not all of those identities have been used, we decided to present all that we have found for future reference.
Some of them we have not proven analytically, but checked numerically for all $N<30$.

Recall that ${\cal M}_N$ is an $N+1\times N+1$  matrix  given by \eqref{eq:M} and
$ {\cal M}^{(a,b)}_N$ is a matrix obtained from ${\cal M}_N$ by deleting the $a^{\text{th}}$ row and the $b^{\text{th}}$ column. It is easy to see that ${\cal M}_{N}^{(1,1)}={\cal M}_{N-1}$.

\subsection*{Determinants with a row/column removed}

 Using \eqref{eq:Iminus} and the fact that $\det{\cal M}_N^{(a,b)}$ is proportional to $\({\cal M}^{-1}\)_{ba}$ it is possible to show that
\begin{align}
\det{\cal M}_N^{(a,b)}=\det{\cal M}_N^{(N+2-b,N+2-a)},\;\;1\le a,b\le N+1.
\end{align}
For any even $N$
\begin{align}
&\det{\cal M}_N^{(1,2)}=-\det{\cal M}_N^{(2,1)},\\
&\det{\cal M}_N^{(a,1)}=(-1)^{a+N/2+1}\det{\cal M}_N^{(N+2-a,1)},\;\;1\le a\le N+1.
\end{align}
For any odd $N$
\begin{align}
&\det{\cal M}_N^{(N+1,1)}=0,\\
&\det{\cal M}_N^{(a,1)}=(-1)^{a+(N+1)/2}\det{\cal M}_N^{(N-a+1,1)},\;\;1\le a\le N
\end{align}

\subsection*{Derivative of a determinant}
For any integer $N$
\beq
\frac{1}{2g}\partial_\theta \det{\cal M}_{N-1}=\det{\cal M}_N^{(2,1)}-\det{\cal M}_N^{(1,2)}.
\eeq

\subsection*{Deformed Bessel functions.}

The ``deformed'' Bessel functions  $I_n^{\theta}$ defined by \eqref{eq:In} satisfy
\beq
I_n^\theta=(-1)^{n+1}I_{-n}^\theta,\ \ \ \ \ \ \ \ I_0^\theta=0.
\label{eq:Iminus}
\eeq
In addition,
\beq
\partial_\theta I_n^{\theta}=2g\left( I_{n-1}^{\theta}- I_{n+1}^{\theta}\right).
\eeq
The first identity is obvious from the definition and the second one is easy to see from the generating function representation \eqref{eq:besselexpansion}.

\end{document}